\documentclass[usegraphicx,useAMS]{mn2e}

\def\arcsecdue {\hbox{arcsec$^{-2}$}}
\def\msun {\ensuremath{M_{\odot} \;}}
\def\lsun {\ensuremath{L_{\odot} \;}}
\def\cmdue {\hbox{cm$^{-2}$ }}

\def\arcsec {\ensuremath{^{\prime\prime}}}
\def\arcmin {\ensuremath{^{\prime}}}
\def\arcdeg {\ensuremath{^{\circ}}}
\def\kms {\hbox{km\,s$^{-1}$}}
\def\yr {\hbox{yr$^{-1}$}}

\def\gtrsim{\mathrel{\hbox{\rlap{\hbox{\lower4pt\hbox{$\sim$}}}\hbox{$>$}}}}
\def\lesssim{\mathrel{\hbox{\rlap{\hbox{\lower4pt\hbox{$\sim$}}}\hbox{$<$}}}}
\def\II {\hbox{{\sc ii}}}
\def\III {\hbox{{\sc iii}}}
\def\yr {\hbox{yr$^{-1}$}}

\def\fe {[Fe/H]}

\def\ml {\ensuremath{M_{HI}/L_B \;}}
\def\hi {\hbox{H{\sc i} }}
\def\hii {\hbox{H{\sc ii} }}

\begin{document}

\title[Evolution of  gas-rich dwarfs in Centaurus A]
{Star formation history and evolution of gas-rich dwarf galaxies
in the Centaurus A group\thanks{Based on observations with: the
NASA/ESA {\em Hubble Space Telescope}, obtained at the Space
Telescope Science Institute, which is operated by the Associations
of Universities for Research in Astronomy (AURA), Inc., under NASA
contract NAS 5-26555; the Australia Telescope Compact Array which
is part of the Australia Telescope, funded by the Commonwealth of
Australia for operation as a National Facility managed by CSIRO.}}

\author[M. Grossi et al.]
{M. Grossi$^{1,2}$, M. J. Disney$^{1}$, B. J .Pritzl$^{3}$, P. M.
Knezek$^{4,5}$\thanks{Visiting Astronomer, Kitt Peak National
Observatory, National Optical Astronomy Observatories, which is
operated by the Association of Universities for Research in
Astronomy, Inc. (AURA) under cooperative agreement with the
National Science Foundation.}, J. S. Gallagher$^{6}$,
\newauthor R. F. Minchin$^{7}$ and K. C. Freeman$^{8}$\\
$^{1}$ School of Physics and Astronomy, Cardiff University, Cardiff, CF24 3YB, UK\\
$^{2}$ Istituto di Fisica dello Spazio Interplanetario INAF-IFSI, Via del Fosso del Cavaliere 100, 00133 Roma, Italy\\
$^{3}$ Macalester College, 1600 Grand Avenue, Saint Paul, MN
55105, USA\\
$^{4}$ National Optical Astronomy Observatory, P.O. Box 26732,
Tucson, AZ 85726, USA\\
$^{5}$ WIYN Consortium, Inc., P.O. Box 26732, Tucson, AZ 85726, USA\\
$^{6}$ Dep. of Astronomy, University of Wisconsin,
Madison, WI 53706-1582, USA\\
$^{7}$ Arecibo Observatory, HC03 Box 53995, PR 00612, USA\\
$^{8}$ Research School of Astronomy and Astrophysics, Mount
Stromlo Oservatory, Cotter Road, Weston, ACT 2611, Australia}

\maketitle
\begin{abstract}

We analyse the properties of three unusual dwarf galaxies in the
Centaurus A group %at a distance of about 4.5 Mpc
discovered with the HIPASS survey. From their optical morphology
they appear to be low surface brightness dwarf spheroidals, yet
they are gas-rich (\ml $>$ 1) with gas-mass-to-stellar light
ratios larger than typical dwarf irregular galaxies. Therefore
these systems appear different from any dwarfs of the Local Group.
They should be favoured hosts for starburst, whereas we find a
faint star formation region in only one object. We have obtained
21-cm data and Hubble Space Telescope photometry in $V$ and $I$
bands, and  have constructed %$I$, $(V - I)$
Colour Magnitude Diagrams (CMDs) to investigate their stellar
populations and to set a constraint on their age. From the
comparison of the observed and model CMDs we infer that all three
galaxies are at least older than 2 Gyr (possibly even as old as 10
Gyr) and remain gas-rich because their star formation rates (SFRs)
have been very low ($\lesssim 10^{-3}$ \msun \yr) throughout. In
such systems, star formation (SF) appears to have been sporadic
and local, though one object (HIPASS J1321--31) has a peculiar red
plume in its CMD suggesting that many of its stars were formed in
a "miniburst" 300 - 500 Myr ago. The question of why there are no
similar  dwarf galaxies in the Local Group remains open. %%%and the
%%%study of similar objects in nearby groups is necessary to widen
%%%the  sample of galaxies with such properties.

\end{abstract}


\section{Introduction}

\subsection{Overview}

Galaxy evolution is related to the rate at which the gas content
is transformed into stars. Along the Hubble sequence this process
appears to be very efficient, and a clear trend holds between gas
content and morphological type with gas-poor early-type galaxies
at one end of the sequence and gas-rich late-type Sd spirals at
the other (Roberts \& Haynes 1994). The gas-mass-to-luminosity
ratio (\ml) is generally taken as a measure of how gas-rich a
galaxy is, with most optically-selected galaxies having $\ml < 1$
 in solar units. However, {\em blind} neutral hydrogen surveys such
as HIPASS (Meyer et al. 2004), HIJASS (Lang et al. 2003) and
HIDEEP (Minchin et al. 2003) have found large numbers of gas-rich
objects where the gas component is dominant compared to the
stellar one (\ml $> 1$). Such objects make up about 30\% of the
detections in these surveys (e.g. Koribalski et al. 2004; Minchin
et al. 2004).

In the optical, gas-rich galaxies often have low surface
brightnesses (LSBs) and low luminosities, but they can also be
found with high star formation rates. Blue compact dwarfs are
gas-rich objects undergoing an intense burst of star formation,
even though their \hi mass-to-light ratios are hardly very high,
being usually in the range 0.1 $\leq$ \ml $\leq$ 2.0 in solar
units. It is not only dwarf galaxies that can be gas-rich, \hi
massive systems with high \ml ratios have been found in HIPASS
(Kilborn et al. 2002). Nevertheless the highest values of \ml are
found in objects that appear in the optical as LSB, low luminosity
dwarf galaxies (van Zee et al. 1997a,b, van Zee 2001a; Warren et
al. 2004) and such systems are the subject of this paper.

We present the properties of three gas-rich dwarf galaxies
($M_{HI} \lesssim 10^7$ \msun) recently discovered at 21-cm in the
Centaurus A group (Banks et al. 1999, Minchin et al. 2003). To all
optical appearances they look like dwarf spheroidals but their
gas-to-stellar ratios (\ml) range from 1.4 to 5 in solar units
which means they have no similar counterparts in the Local Group
(LG). They are puzzling because, in principle, they represent an
ideal environment for SF, given their large gas fractions and
their location in a fairly populated group such as Centaurus A.
Yet their \ml ratios indicate that a large fraction of the
galactic gas reservoir has not been converted into stars yet.

Are we finding a new class of galaxies? Are they young? Are there
any common properties with the well studied dwarf systems of the
LG? Do these gas-rich dwarfs show star formation histories (SFHs)
similar among themselves, or are they different one from the
other? The aim of this analysis is to try to understand some of
these issues.

%%%%As a first step we need to identify their different stellar
%%%%populations providing information on the age of these galaxies.
One possible explanation is that they represent a local population
of {\em young} ($<$ 1 Gyr) galaxies, which are gas-rich because
they have not existed long enough to transform their gas into
stars. The presence of  nearby young systems has been under debate
for years until the recent discovery that the  best candidate for
a nearby young galaxy, the blue compact dwarf (BCD) galaxy I Zw 18
(12.6 Mpc $< d <$ 15 Mpc), does not contain a population of stars
older than $\sim$ 500 Myr (Izotov \& Thuan 2004). Being at a
distance of less than 5 Mpc, the Centaurus A dwarfs may represent
ideal candidate young galaxies to study.

Alternatively, these systems could be {\em old} galaxies, that
formed their first stellar populations several Gyr ago, but have
evolved at a slower rate compared to "normal" Local Group galaxies
i.e. they are in some sense {\em ``retarded''}. %%%It their Star formation
%%%rate (SFR)  has been continuous, then it must have been highly
%%%suppressed during their evolution, and
If this is the case, then what prevents stars from forming in such
objects? %%%What are the processes that can inhibit the conversion of
%%%gas into stars?
If one could answer  this question   we might learn something
about galaxy evolution in general.

In order to investigate the Star Formation Histories (SFHs) of
these dwarf galaxies we have obtained Hubble Space Telescope (HST)
images in $V$ and $I$ from which we have built CMDs. These are
compared to 21-cm maps obtained with the Australia Telescope
Compact Array (ATCA).

This paper will proceed as follows: first an overview of the
Centaurus A group is given in the remaining part of this
introductory section (\S 1.2). In section 2 we describe both the
21-cm and the optical data we have obtained. In sections 3, 4, 5
we analyse the three dwarfs, HIPASS J1337--39, HIDEEP J1337--3320
and HIPASS J1321--31 individually. Finally our results will be
summarised  in \S 6. In \S 7 we present our  conclusions.

\subsection{The Centaurus A group}

The Centaurus A group is a heterogeneous assembly of early to
late-type galaxies at a distance of between 3.5 and 4.5 Mpc,
covering about 1000 square degrees of sky.  The two largest
galaxies in the group are NGC 5128 (Cen A), the nearest early-type
giant and a very active radio galaxy, and the spiral NGC 5236 (M
83)

The distance of the group was originally set by de Vaucouleurs at
$d =$ 4.0 Mpc (1975). Soria et al. (1996) placed NGC 5128 at 3.6
$\pm$ 0.2 Mpc using the tip of the red giant branch (TRGB) while,
also from the TRGB, Harris et al. (1998) found a distance of 3.9
$\pm$ 0.2 Mpc.

NGC 5236 appears to be more distant than NGC 5128: Thim et al.
(2003) derived $d_{M83} = 4.5 \pm 0.3$ Mpc from Cepheid
observations. Its close neighbour, NGC 5253 (angular separation
110 arcmin), is at $d = 4.0 \pm 0.3$ Mpc from Cepheid and SN Ia
(Saha et al. 1995) measurements.

In recent years, the number of smaller galaxies associated with
the group has increased greatly.  C\^ot\'e et al. (1997) added 20
dwarf galaxies %% to the group
(mainly gas-rich dwarf irregulars) of which 2 were newly
identified. Four gas-rich neighbours were found around M 83 in
this survey, but none near NGC 5128, presumably because its hot
gaseous halo make Cen A a very hard environment for the
''survival'' of gas-rich satellites.

An \hi survey of the group by Banks et al. (1999) added another
ten members, five of which were previously  known galaxies with
wrong or no redshift measurements, and a deeper, smaller-area \hi
survey, HIDEEP  (Minchin et al. 2003), added another new dwarf.
All of the new six detections appeared to have  optical
counterparts, although very faint ($M \sim -11$) and LSB ($\mu_0^B
\gtrsim 24$ mag \arcsecdue).

Another optical survey covering the same area as  C\^ot\'e and
collaborators, aimed at finding exclusively dwarf elliptical
candidates, added 13 new dSph members to the group (Jerjen et al.
2000a, 2000b).  The survey led also to the addition of one more
dIrr galaxies (AM1318-444).

Seven more members have been included in
%%added to
the group by Karachentsev et al. (2002) with a survey of nearby
dwarf galaxy candidates which is part of the Hubble Space
Telescope snapshot (SNAP) program. The majority of the targets
were taken from the results of an all-sky search for nearby dwarf
galaxies, based on the POSS II and ESO/SERC plates and were mainly
dwarf spheroidal, LSB galaxies.

To date, 62 candidate members of the group have been classified,
and a measure of the distance is  available for about thirty
objects (Karachentsev 2005).

As regards the inner structure of the Centaurus A group,  de
Vaucouleurs (1975), Tully (1987), Van den Bergh (2000a), consider
that all the brightest members form a single group, while
Karachentsev (1996) argues that  the two massive galaxies, NGC
5128 (CenA) and NGC 5236 (M 83) are the centres of two separate
subgroups (Karachentsev et al. 2002), at a distance $d_{CenA} =
3.66 \pm 0.07$ Mpc and $d_{M83} = 4.57 \pm 0.05$ Mpc respectively.
Moreover the M 83 group is moving away from Cen A at a relative
radial velocity of 55 \kms (Karachentsev et al. 2002).

Whether the two groups are separated or not partly depends on the
assumed distance of NGC 5128, which is found to range in the
literature between 3.6 and 4.0 Mpc. However, Karachentsev and
collaborators find a clear distinction between objects with
average distance around 3.6 Mpc and those whose average distance
is 4.5 Mpc: a result that strengthens the two subgroups
interpretation.

To conclude, the Centaurus A complex of galaxies appears as a very
rich environment within the Local Volume, with a variety of
members spanning different morphological types, hardly found in
other nearby groups. Sculptor, for instance, lacks both early-type
members and bright Milky-Way-like spiral galaxies, while the M 81
group has about 28 members (with only one large spiral and 23
dwarf candidate members) but in contrast with Centaurus A, it
is a very compact group. %%%Its core galaxies
%%%- the large spiral M 81, the peculiar galaxies M 82 and NGC 3077, and
%%%the small spiral galaxy NGC 2976
%%%%are strongly interacting as it is
%%%%indicated by the big \hi structure which surrounds them (van Driel
%%%%et al. 1998; Boyce et al. 2001).

\section{Observations and data reduction}

\begin{figure*}
\begin{minipage}{126mm}
\includegraphics[width=63mm]{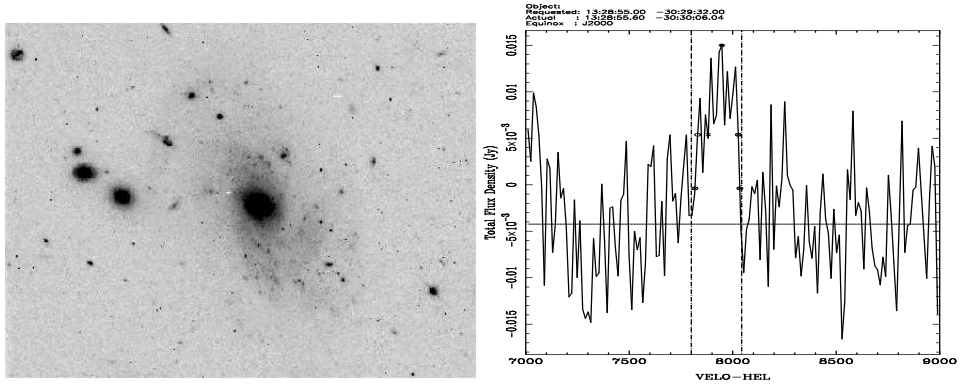}
\includegraphics[width=63mm,height=50mm]{Figure1b.eps}
%\vspace{63mm}
\caption{HIPASS J1328-31. {\em Left:} WFPC2 image taken with the
filter F814W. {\em Right:} 21-cm spectrum taken from the HIDEEP
survey (Minchin et al. 2003).} \label{1328-30}
\end{minipage}
\end{figure*}

Four among the six new gas-rich dwarf galaxies from HIPASS and
HIDEEP were chosen to be followed up with the HST's Wide Field
Planetary Camera 2 (WFPC2) and at 21-cm with ATCA. The remaining
two  (HIPASS J1348-37, HIPASS J1351-47) have both high \ml ratios,
respectively 2.7 and 3.9, but they were not included in the list
because they had been already observed with the WFPC2 as part of
the SNAP survey.

\subsection{ATCA \hi observations }

Observations were carried out with ATCA in a 750-m configuration
in May 2001 and in a 1.5-km configuration in October 2001.  Both
sets of observations had an integration time of 12 hours per
source and used a bandwidth of 4 MHz, divided into 1024 channels
to give a velocity resolution of 0.8 \kms.  PKS 1934-638 was used
as the flux calibrator and PKS 1320-1446 as the phase calibrator.

The first run gave short-baseline data that were then  combined
with the long-baseline ones %from the second run
to give a good compromise between sensitivity and resolution.  The
first data set also
%enable a check to be made
%enabled to make a
allowed us to check  the reality of the sources; one of the
objects (HIPASS J1328-30) turned out to be a more distant
uncatalogued LSB galaxy at $v_{opt} \sim 8100$ \kms %%%(Freeman, {\em
%%%private communication})
which overlaps a local high velocity cloud
with $v_{helio} \sim 200$ \kms. This was confirmed by the WFPC2
image of this object (Fig. \ref{1328-30}). The galaxy shows a
compact fairly bright core, with an extended LSB disc with spiral
arms. The corresponding dereddened total magnitude is $m_{I,tot} =
11.65 \pm 0.03$ mag.  This galaxy was therefore not observed in
the second run.

Data reduction was carried out in {\scshape miriad}.
%%, with the data being continuum subtracted in the U-V plane.
For HIPASS J1321--31 and HIDEEP J1337--3320 a robust setting of
0.5 was used, giving a lower noise level at the expense of some
angular resolution.  This gave a noise of 3 mJy per channel with
beam-sizes $\sim$ 57\arcsec $\times$ 29\arcsec (HIPASS J1321--31)
and $\sim$ 59\arcsec $\times$ 32\arcsec (HIDEEP J1337--3320).  For
HIPASS J1337--39, uniform weighting was used to obtain the best
angular resolution. This gave a noise of 5 mJy per channel and a
beam-size  $\sim$ 31\arcsec $\times$ 24\arcsec.

ATCA and single dish fluxes obtained by the HIPASS survey with the
Parkes telescope have been compared to check if a component of the
flux was missing in the interferometric detections, but we found
that variations are in the order of the 10\%, consistent with the
errors in HIPASS fluxes.

\begin{table*}
\begin{minipage}{126mm}
\caption{Observed properties of the three Centaurus A dwarfs}
\label{tabprop}
\begin{tabular}{lcccccccc}
\hline \hline SOURCE  & RA & DEC & $v_{helio}$ & r & $m_B$ &
$M_{B}$ &  $M_{HI}$
&$M_{HI}/L_B $ \\
 & J2000 & J2000 & (\kms) & (kpc) & & & $10^7 M_{\odot}$  & ($M_{\odot}/L_{\odot}$) \\
\hline \hline
HIPASS J1337--39  & 13:37:26 & -39:52:15 &  491 & 0.40 & 16.1 & -12.6 & 3.9   &   2.1  \\
HIDEEP J1337--3320 & 13:37:01 & -33:21:47 &  590 & 0.22 & 17.5 & -10.9 & 0.5  &  1.4 \\
HIPASS J1321--31  & 13:21:06 & -31:32:25 & 572   & 0.76 & 17.1 & -11.7 & 3.7  & 4.9 \\
\hline \hline
\end{tabular}
\end{minipage}
\end{table*}

The main optical and \hi  properties of the three HIPASS dwarfs
are shown in Table \ref{tabprop}.

\subsection{WFPC2 optical observations}

The three targets were followed up with WFPC2 in June 2001 through
the filters F555W and F814W.  These two filters were chosen for
ease of transformation to the Johnson/Cousin $V$ and $I$ system
which is the standard for observing both old and young stellar
populations. To get the largest spatial coverage possible the
galaxies were centred in the WF3 chip of the camera, with a
spatial sampling of 0\arcsec.1 per pixel and a field of view of
80\arcsec $\times$ 80\arcsec. Four orbits were assigned to each
target for a total of 5000 s (F555W) and 5200 s (F814W).

The frames were debiased, zero-corrected, dark-subtracted and
flat-fielded by the Space Telescope Science Institute (STScI)
pipeline process before being made available. They were then
combined with the task {\scshape crrej} in {\scshape
iraf}\footnote{{\scshape iraf} is distributed by the National
Optical Astronomy Observatories, which are operated by the
Association of Universities for Research in Astronomy, Inc., under
contract to the National Science Foundation.} to remove
contamination by cosmic rays on the frames. The photometry was
done using the {\scshape iraf} version of the package {\scshape
daophot} and {\scshape allframe} (Stetson 1987). A parallel
analysis of the data (Pritzl et al. 2003) has been performed by
our collaborators at the National Optical Astronomy Observatory
(NOAO) using the stand-alone version of {\scshape daophot}. A
preliminary selection of stars with $S/N >$ 3 was performed with
the automatic star-finding algorithm {\scshape daofind} and their
magnitudes obtained with the task {\scshape phot} using an
aperture radius of 2.0 pixels. An empirical point-spread function
was built with the task PSF by selecting several isolated and
bright stars in each frame containing the galaxies. The
instrumental magnitudes were measured with the {\scshape allstar}
algorithm (Stetson 1994) by fitting the point-spread function to
the brightness profile of any given object.

When measuring the photometry of stars in a crowded field some
mistakes and wrong detections are inevitable. Defective pixels,
cosmic rays, background objects, and stellar blends can
''pollute'' the output of the measured photometry performed by
{\scshape allstar}. However the routine produces indices that give
an indication of the reliability of the fit and provide a tool to
flag and reject dubious detections.

These indices are: $\chi$,  the square root of the standard
$\chi^2$ goodness-of-fit index and represents a dimensionless
measure of the agreement between the measured brightness profile
and the model PSF for any given stellar detection.
%%%In principle $\chi \approx 1$, but significantly larger values may
%%%indicate non-stellar objects or profiles corrupted by image
%%%defects (Stetson, Bruntt, \& Grundhal 2003). At increasingly faint
%%%magnitudes $\chi$ tends to converge around 1 because the
%%%background noise dominates over any intrinsic abnormalities in the
%%%objects' brightness profile. For the brightest stars $\chi$ tends
%%%to assume values much greater than 1.
 The residuals of the fitting, $\sigma$, that are representative
of the errors on the derived instrumental magnitudes. Lastly,
$sharp$ is a first order estimate of the intrinsic angular radius
of a resolved source: if the PSF has a characteristic radius
$s_{PSF}$ and the measured image profile has a radius $s_{obs}$,
$sharp^2 = |s^2_{obs} - s^2_{PSF}|^2$. $Sharp$ values should be
tightly clustered around zero for bright stars, while the
deviation from zero increases for fainter stars.
%%%%When $sharp$ is
%%%%positive but too large, it can be taken as an indication of a
%%%%resolved non-stellar object. If $s_{obs} < s_{PSF}$, sharp assumes
%%%%a negative value, and when it is too small, it is probably related
%%%%to an image blemish or a cosmic ray.

\begin{figure}
      \includegraphics[width=84mm]{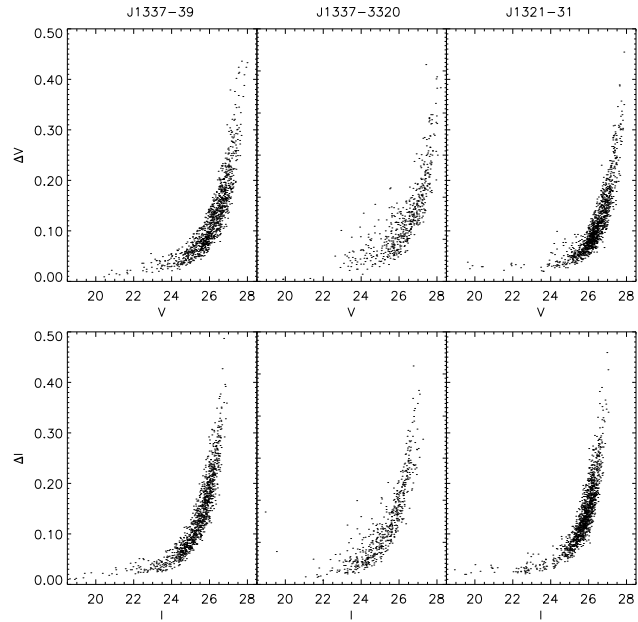}
      %%{chishperr39.ps}
%\vspace{84mm}
      \caption{HIPASS J1337--39: Plot of photometric errors  vs.
apparent visual magnitudes in both $V$ ({\em upper panel}) and $I$
({\em lower panel}) bands after applying the rejection criteria
and transforming to the Johnson system.}
   \label{shperr}
      \end{figure}

Therefore, to minimise the number of false detections we have
included only stars with $-0.5 < sharp < 0.5$.  For stars with
poor photometry (errors $> 0.1$ mag), an additional criterion of
$\chi < 2$ was applied.  This choice should provide a
representative sample of detections for each dwarf, excluding both
objects smaller than the size of a star, such as cosmic rays or
image defects, and extended objects like galaxies, \hii regions,
stellar blends and any remaining blemishes. We have shown in Fig.
\ref{shperr} the plots of $V$ and $I$ magnitudes versus their
relative photometric errors calculated by {\scshape allstar},
after having applied the rejection criteria.

      \begin{figure}
      \includegraphics[width=84mm]{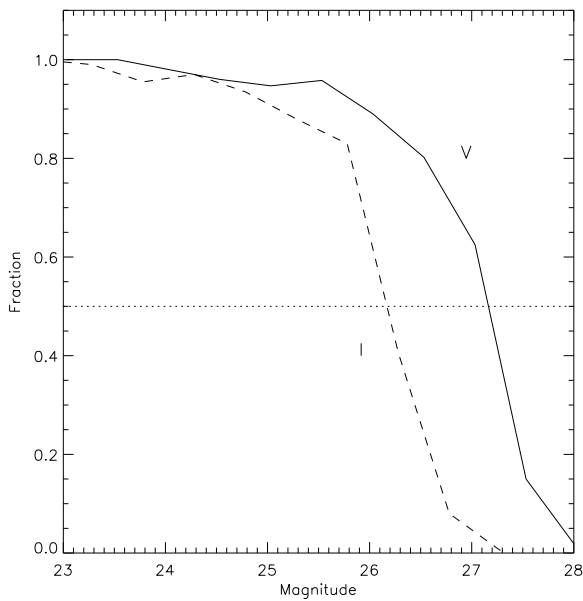}
      \caption{Photometry completeness expressed in percentage of recovered stars. Since we do not
      find large variations between the three fields, for the sake of clarity here we show
      only the case of HIPASS J1337--39. The horizontal dotted line represents the 50\% level.}
   \label{completeness}
      \end{figure}

The transformation from the F555W and F814W magnitudes to the
Johnson-Cousin system was performed using the iterative procedure
described in Holtzmann et al.(1995a) and the colour terms given in
their Table 7. %%%%Using bright and isolated stars in each frame
%%%%containing the galaxies,
We determined the aperture corrections from the PSF fitting
photometry to the 5 pixels aperture considered in Holtzmann et al.
(1995a). Further corrections were applied for the exposure time,
the gain factors and the charge transfer efficiency (CTE), using
the equations
derived by Dolphin (2000a).   %%%In particular, we have
%%%used the following relation between ALLSTAR magnitudes $v$ and $i$
%%%and the standard magnitudes $V$ and $I$
%%%
%%%\[ V = v + 2.5 \log \; t_{exp} + A_V - 0.052 \times (V - I) - 0.027 \times (V-I)^2 \]
%%%
%%%\[ I= i +2.5 \log \; t_{exp} + A_I - 0.062 \times (V - I) - 0.025 \times (V-I)^2 \]
%%%
%%%where $A_V = 22.02$ and  $A_I=21.17$ are the sum of the zero
%%%point, minus the gain factor (0.75 mag for WF3 and the 7$e^-$
%%%gain), and the aperture correction term.
%%%%Finally CTE corrections have been introduced using the equations
%%%%derived by Dolphin (2000a).
The final lists of stars were matched in both filters assuming a
matching radius of 1 pixel (0.1\arcsec). %%%%The resulting colour
%%%%magnitude diagrams (CMDs) for the three dwarfs are shown and
%%%%discussed in the next chapters.

The completeness or our photometry has been tested by adding
artificial stars of known $I$ and $V$ magnitudes to the original
frames using the {\scshape addstar} algorithm of {\scshape
daophot}. For each 0.5 bin of magnitude we have added 400 stars.
To avoid overcrowding effects we have created four images per each
bin and included 100 uniformly distributed artificial stars in
each of them. Each frame was then processed in exactly the same
way as the original images. The resulting photometry has been
matched with the original list of stars and magnitudes created
with {\scshape addstar}. The percentage of recovered stars (shown
in Fig. \ref{completeness}) indicate that completeness is around
50\% for $V$ = 27.2 mag and  $I$ = 26.2 mag for HIPASS J1337--39,
and is approximately the same for the other galaxies.

\begin{figure*}
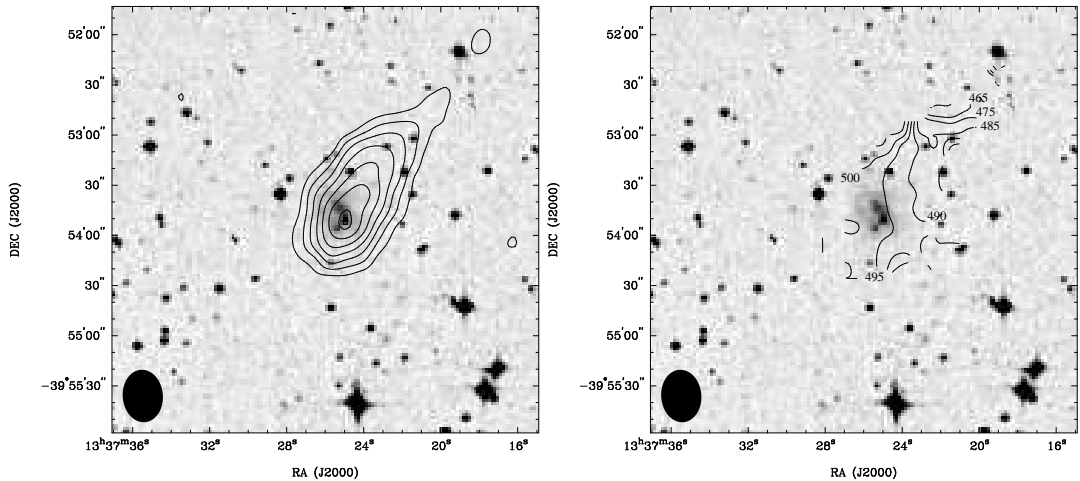

\begin{minipage}{156mm}
\includegraphics[width=63mm,angle=-90]{Figure4a.eps}
\includegraphics[width=63mm,angle=-90]{Figure4b.eps}
%\vspace{63mm}
    \caption{{\em Left:} \hi density contour maps of HIPASS J1337--39  overlaid on the Digital
     Sky survey field including the galaxy. The contour levels are:
2.5, 5.0, 7.5, 10, 12.5, 15, 17.5 $\times 10^{20} $ cm$^{-2}$.
 {\em Right:} The \hi velocity field: contours are in the range between 465
\kms $\;$and 500 \kms. The beam size is shown at the bottom left
corner of the image.}
    \label{39HI}
\end{minipage}
\end{figure*}

\subsection{H$\alpha$ observations}

Direct H$\alpha$ imaging observations were made at the WIYN
Observatory\footnote{The WIYN Observatory is a joint facility of
the University of Wisconsin-Madison, Indiana University, Yale
University, and the National Optical Astronomy Observatories.} 3.5
m telescope. The data were taken on 29 and 30 May 2002 with
MiniMo, which has two SITe 4096 x 2048 thinned CCDs. The pixel
scale is 0.14 arcsec/pixel and the field of view is 9.6
arcminutes. Broadband $R$ images were made using a Harris filter,
and narrowband H$\alpha$ imaging was done using a filter centered
at 6570 $\AA$ and 73 $\AA$ wide. For each dwarf galaxy the
integrations time were 3 minutes in $R$, and 30 minutes (3 images
of 10 minutes each) in H$\alpha$.  The Landolt standard field
PG1323-086 (Landolt 1992) was observed each night in both filters,
along with the spectrophotometric standard Cyg OB2 No. 9 (Massey
et al. 1988).

Data processing was done using the {\scshape mscred} reduction
tasks in {\scshape iraf}. In order to estimate the H$\alpha$ flux
for each galaxy, the observations of the spectrophotometric
standard Cyg OB2 No. 9 were used.  Only the galaxy HIPASS J1337-39
was detected in emission.
%The flux measured is $6.9 \times 10(-14)$ ergs cm$^{-2}$
%sec$^{-1}$.
%Because only one spectrophotometric standard was measured in
%H$\alpha$, we estimate that the flux is uncertain at the 50\%
%level.
For HIDEEP J1337-33 and HIPASS J1321-31, the flux limit is
derived by measuring the flux in the faintest detected stars in
each image. For these galaxies, we estimate that the H$\alpha$
flux is less than $7 \times 10^{-16}$ ergs cm$^{-2}$ sec$^{-1}$.

Then we have observed HIPASS J1337-39 with the Double Beam
Spectrograph (DBS) on the 2.3 m telescope at the Siding Spring
Observatory\footnote{The 2.3 meter Telescope is run by the
Australian National University as part of the Research School of
Astronomy and Astrophysics.} to study the chemical abundance of
the \hii region and to obtain an independent measure of the
metallicity. The DBS provides a spectral coverage of 3600 - 7000
\AA.

Two exposures of 1337-39, each 2000 s long, were taken with both
the blue and red grism. We performed the wavelength calibration
using Neon - Argon lamps with exposures taken before and after the
object observations. After the flux calibration (performed with
the spectrophotometric standard EG21 and LTT3864; Hamuy et al.
1992, 1994), one-dimensional spectra were extracted from the
two-dimensional images by summing up all the pixels enclosed in
the profiles of the H$\alpha$ line. The background and sky lines
were subtracted by averaging a region on either side of the
extraction window. We have then identified the main emission lines
and measured their fluxes. The data were reduced with the
{\scshape iraf} package. The observed line intensities were
corrected for interstellar extinction by comparing the observed
and the expected hydrogen line ratios. Assuming the theoretical
Balmer line value for case $B$ recombination (Osterbrock 1989), we
have determined the reddening constant $C(H\beta) = (\log
R_0/R)/\phi(\lambda)$ where $R_0 = I_{\lambda,0}/I_{H\beta,0}$ is
the theoretical H line ratios relative to H$\beta$, $R$ is
 the observed ratio, $\phi(\lambda) = f(\lambda)-f(H\beta)$ is the extinction function relative to the H$\beta$ line, where $f(\lambda)
 = < A(\lambda)/A(V) >$. For $A(\lambda)$ we have assumed the reddening law of Cardelli et al. (1989), with $R_V = 3.1$.

\section{HIPASS J1337--39}

Among the three dwarfs, J1337--39 is the only galaxy with ongoing
star formation activity. With an \hi mass of 3.9 $\times 10^7$
\msun and a total apparent magnitude $m_B = 16.1$ (Doyle et al.
2005), it has a gas-mass-to-stellar ratio \ml $\simeq 2$ in solar
units.

\subsection{The neutral gas content}

The neutral gas in J1337--39 (Fig. \ref{39HI}) shows a smooth
distribution with a one-sided extension to the north west
direction out to a projected radius of 1\arcmin 30\arcsec, or
$\sim$ 2 kpc at the assumed distance of 4.8 Mpc (see \S 3.5).
J1337--39 has the highest average \hi column density of the three
dwarfs, and gas can be traced out to an \hi column density of $2.5
\times 10^{20}$ \cmdue. The overall high column density may
explain why this is the only galaxy with ongoing star formation.
The density peak is at $\sim 2 \times 10^{21}$ \cmdue in
correspondence with the star forming regions and slightly offset
from the optical centre of the galaxy (see Fig. \ref{39HI}).

The velocity field is shown in the right panel of Fig. \ref{39HI}.
There is evidence of a small velocity gradient throughout the
galaxy; the disc appears to be slowly rotating although
the rotation is not very sharply defined. The velocity contours
become rather irregular in the outer edge of the galaxy, in
correspondence with the edge of the \hi ``tail'', while around the
main optical body the velocity field looks fairly regular.

From the 21-cm integrated flux we derive an \hi mass $M_{HI} = 3.9
\times 10^7$ $M_{\odot}$, at a distance of 4.8 Mpc. The \hi
profile has a characteristic width $\Delta v_{20\%} \sim 50$ \kms.
A rough estimate of the dynamical mass (without any
inclination-angle correction) gives $M_{dyn} = V^2_R \times
R_{HI}/G \sim 4 \times 10^8$ \msun, and the ratio $M_{HI}/M_{dyn}
\sim 0.1$ is similar to the sample of gas-rich dwarfs studied by
van Zee et al. (1997a) and to what is found for ESO0215-G?009
($M_{HI}/M_{dyn} = 0.11$), one of the most gas-rich galaxies ever
detected with \ml = 22 (Warren et al. 2004). Roberts \& Haynes
(1994) determined for their sample of Sm/Im galaxies a median
value of 0.15.

The role of chaotic motions is in general more important in low
mass galaxies (Young \& Lo, 1996, Lo \& Sargent 1993). As an
estimate of the dispersion of the gas, in order to calculate the
balance between rotation and chaotic motions, we have calculated
the second moment with MIRIAD, defined as

\[ \sigma = M_2 = \sqrt \frac{ \int I(v) (v-M_1)^2 dv}{\int I(v) dv} \]

which corresponds to the intensity weighted velocity dispersion
squared, where $M_1$ is the intensity weighted  velocity defined
as $\int I(v) v dv / \int I(v) dv$.

For HIPASS J1337--39 the resulting velocity dispersion is
$\sigma_v \simeq 10$ \kms. The velocity resolution of our data is
$\sim$ 1 \kms, so we are not measuring an instrumental effect.
Although the velocity dispersion is comparable to the rotational
velocity $V_R = 25$ \kms  (without inclination-angle corrections),
chaotic motions appear not to be dominant for this dwarf.

Finally we have inspected the immediate environment of the galaxy
to search for possible companions within the ATCA primary beam
($\sim$ 30\arcmin), but we have not detected any emission from
other sources.

\subsection{Optical properties and the oxygen abundance}

\begin{figure}
\includegraphics[width=84mm]{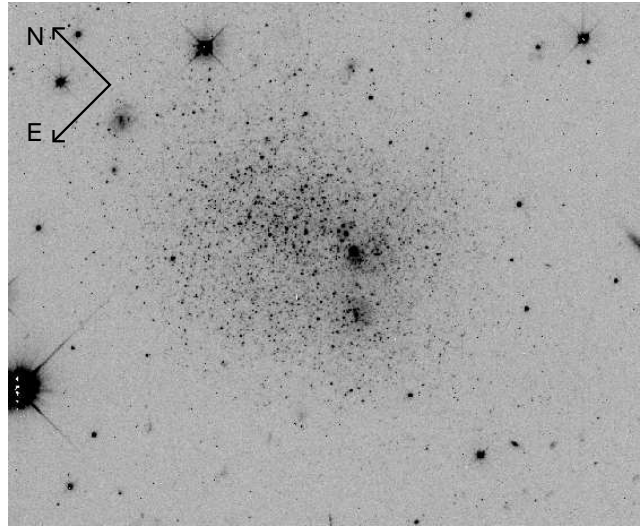}
%\vspace{84mm}
    \caption{HIPASS J1337--39: The WFPC2 image of the galaxy taken with
the filter F555W. The size of the frame is roughly 80\arcsec.
North is on the upper-left corner, and East on the lower-left one.
Two nearby \hii regions are clearly visible in the southern area
of the galaxy.}
    \label{39HST}
\end{figure}

The bulk of the stellar distribution (Fig. \ref{39HST}) shows an
overall spherical symmetry, similar to that found in dSph
galaxies. Outside the main spherical distribution, stars probably
belonging to the dwarf extend north-south along the diagonal of
the chip to the edges of the frame.

Star formation activity in this galaxy  has been confirmed by
H$\alpha$ observations taken with the WIYN telescope and two
nearby \hii regions can be identified also from the WFPC2 image
(Fig. \ref{39HST}), slightly offset from the optical centre.

The symmetric stellar distribution of J1337--39 resembles that of
dwarf spheroidals, while the ongoing SF activity and the large gas
fraction is more typical of dIrr galaxies. %%%%% The optical appearance of

Once H$\alpha$ emission had been detected, observations with the
Double Beam Spectrograph (DBS) on the 2.3 m telescope at the
Siding Spring Observatory were taken to obtain an independent
measure of the current metal abundance of the ISM. This allows us
to break the age-metallicity degeneracy that represents the main
difficulty in the study of the stellar populations from CMDs.

\begin{table}
\begin{center}
\caption{Line flux ratios relative to H$\beta$. We also show the
value of the reddening constant, $C(H\beta$) and  the H$\alpha$,
H$\beta$ fluxes.} \label{line-intensities}
\begin{tabular}{cc}
  \hline \hline
  % after \\: \hline or \cline{col1-col2} \cline{col3-col4} ...
  Ionic species & $F(\lambda)/F(H\beta)$ \\
  \hline \hline
$[O\II]$  & 0.20 $\pm$ 0.06 \\ %0.2611 \\
H$\beta$   & 1. $\pm$ 0.05\\
$[O\III]$(5007$+$4959)  & 3.10 $\pm$ 0.28 \\ %3.6183 \\
H$\alpha$   & 2.87  $\pm$ 0.20\\ %%2.8702 \\
%$[S\II]$ 6717   & 0.07 $\pm$ 0.05\\ %%0.0668 \\
%$[S\II]$ 6731  & 0.05  $\pm$ 0.03\\ %%0.0529 \\
%&  \\
 \hline \hline
%&  \\
$C(H\beta) $ & 0.32  \\
$ F(H\beta)$/ergs s$^{-1}$ cm$^{-2}$  & 1.9 $\cdot 10^{-14}$ \\
$ F(H\alpha)$/ergs s$^{-1}$ cm$^{-2}$ & 5.5 $\cdot 10^{-14}$ \\
%%$^a$ In units ergs s$^{-1}$ cm$^{-2}$ & \\
  \hline \hline
\end{tabular}
\end{center}
\end{table}

We detected a number of lines, including H$\alpha$, H$\beta$,
O{\sc ii} and O{\sc iii} (see Table 2), allowing us to determine
the oxygen abundance via the $R_{23}$ indicator (Pagel et al.
1979; Skillman 1989) as 12 + $\log$(O/H) = 7.25 $\pm$ 0.1, or
[O/H] $\simeq$ 1/30 solar\footnote{Given the solar oxygen
metallicity $\log$(O/H)$_{\odot} + 12= 8.83$ (Grevese \& Sauval
1998).}. As a comparison, the abundance of I Zw 18 is 12 +
$\log$(O/H) = 7.2 (Izotov \& Thuan 1999), therefore this galaxy
appears to have a very low metal abundance. Using the relation
between iron and oxygen abundances of Mateo (1998), we obtain
[Fe/H]= $-1.95 \pm 0.2$.

Since the metallicity is a measure of the degree of evolution of a
system, the low abundance that we have derived suggests that
J1337--39 is a poorly evolved system. To finally establish whether
this is the consequence of an overall young age or not, we need to
analyse the results of the resolved stellar photometry.
%% to constrain the age of the galaxy.

\begin{figure}
%\begin{figure*}
%\begin{minipage}{156mm}
\includegraphics[width=87mm]{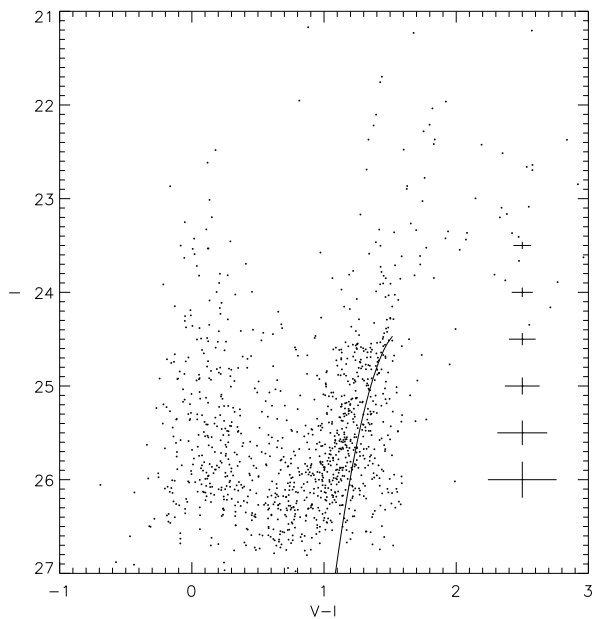}
%\includegraphics[width=63mm]{Figure6b.eps}
%\vspace{63mm}
    \caption{The $(V-I)$, $I$ color magnitude diagram of
HIPASS J1337­-39. As a comparison, we have overlaid  the fiducial
RGB of the galactic globular cluster M 15 ([Fe/H]=$-2.17$).
Photometric errors, obtained from the point spread function (PSF)
fitting, are plotted on the right of the figure.}
%%%{\em Right}: The
%%%CMD of the field region. This diagram has been constructed by
%%%using only the halves of the chips (WF2, WF4) not adjacent to the
%%%WF3 where the galaxy was positioned. This choice reduces the
%%%possible contamination of the field by stars belonging to the
%%%dwarf.}
    \label{39CMD}
%\end{minipage}
%\end{figure*}
\end{figure}

\subsection{The morphology of the CMD}

The CMD of J1337-39 is displayed in the left panel of Fig.
\ref{39CMD}. The right panel  shows the CMD of the two remaining
chips of the WFPC2 where the galaxy was not included that have
been used to estimate the contamination of the field by foreground
stars. The diagram clearly shows both a blue young stellar
population with colours $V - I < 0.7$, {\em the blue plume},
including main sequence (MS) and bright blue loop (BL) stars, and
also a larger number of red stars at colours $0.7 < V - I < 1.5$
which clearly indicates the presence of a  red giant branch (RGB)
population. There is also evidence of a very narrow ``red tail''
elongated for roughly one magnitude above the tip of the RGB with
colours 1.2 $< V - I <$ 1.6, which would suggest the presence of a
metal-poor intermediate-age asymptotic giant branch (AGB)
population. The evidence of a well-populated RGB here means that
this galaxy must be at least older than 1 Gyr; HIPASS J1337--39 is
{\em not} a {\em young} galaxy.

There are several age indicators in a CMD to set constraints on
the age of the galaxy, but with this data set we can only detect
stars which are roughly two magnitudes fainter than the tip of the
RGB (TRGB), thus we are missing the oldest age indicators, such as
the red clump and the horizontal branch stars. Intermediate age
AGB and RGB stars are the oldest stellar population we can detect.

\subsection{The RGB population}

\subsubsection{Deriving the distance to HIPASS J1337--39}

\begin{figure}
\includegraphics[width=74mm]{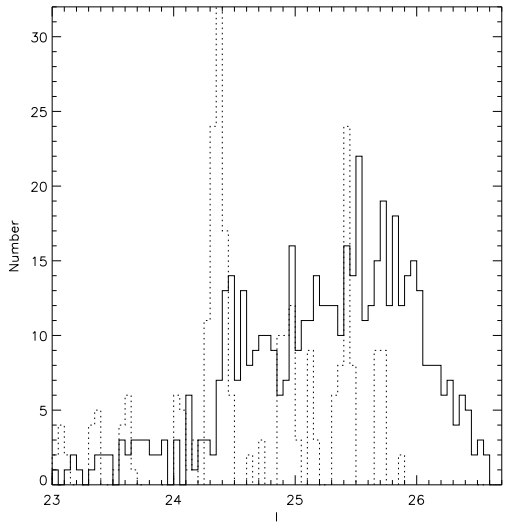}
%\vspace{84mm}
    \caption{The luminosity function of the RGB of HIPASS J1337--39. The
output of the edge detection Sobel filter is overlayed (dotted
line). After having applied the filter we found  the tip of the
RGB at $I_0$ = 24.37 $\pm$ 0.11.}
    \label{lumfunc39}
\end{figure}

The RGB can be seen as a high  concentration of stars at $I >
24.5$ and $0.7 < V - I < 1.5$.  The absolute $I$-band magnitude of
the tip of the RGB is not affected by  metallicity and age
variations (Iben \& Renzini 1983), therefore it has been
introduced as a  robust distance indicator for resolved stellar
systems (Da Costa \& Armandroff 1990, Lee et al. 1993). To
determine this parameter we have built the $I$ luminosity function
of stars with colours $0.7 < (V-I) < 1.7$ (Fig. \ref{lumfunc39}).
An edge detection Sobel filter (Madore \& Freeman 1995) has been
applied to the function. This produces a sharp peak when a sudden
drop in the original function occurs, allowing the TRGB to be
located. A cut-off in the RGB luminosity is found at $I_0 = 24.37
\pm 0.11$, where the magnitude has been corrected for the galactic
extinction using $E_{B-V} = 0.07 \pm 0.01$ and $A_I = 1.85 \times
E_{B-V} = 0.14 \pm 0.02$ (Schlegel et al. 1998). Assuming the TRGB
to be $M_I = -4.05$, the distance modulus is $(m - M)_0 = 28.42
\pm 0.11$, giving a distance $D = 4.8 \pm 0.2$ Mpc. This value
places J1337--39 in the region of M 83 (4.5 Mpc) although it is at
a projected distance of 10\arcdeg\ ($\sim$ 850 kpc at 4.8 Mpc)
from the spiral.

\subsubsection{A constraint on metallicity and age from the RGB}

We can use the RGB to inspect the abundance of the red giant stars
(the oldest population of stars that we can detect). The simplest
method is to compare the colour of the RGB of this dwarf with that
of well-known metal-poor galactic globular clusters (GGCs) (Da
Costa \& Armandroff 1990). In Fig. \ref{39CMD} the fiducial RGB of
M 15, one of the most metal-poor GGCs with \fe = $-$2.17, is
overlaid on the CMD of J1337--39. Most of the RGB stars of
J1337--39 extend blueward of the fiducial line representing the
globular cluster, which suggests an even lower metallicity. We can
estimate the metallicity of the stars in the RGB using the
relation of Lee et al. (1993), even though this technique  relies
upon the hypothesis that the dominant population of the RGB is
several Gyrs old (Population II stars). The mean of the
distribution of the stars in the RGB gives a  de-reddened colour
at $M_{I, 0} = -3.5$, $(V - I)_{-3.5} = 1.14 \pm 0.13$, which
corresponds to  an abundance \fe = $-2.6 \pm 0.6$, placing it
among the lowest metallicity nearby dwarf galaxies.  The error on
the photometry at that magnitude is rather large ($\Delta M =
0.15$) and produces a large uncertainty on the metallicity.  Note
also that  Lee et al. relation  in this case is used in
extrapolation, since the least metallic globular cluster for which
it has been determined  is M 15 at \fe = $-$2.17. In any case, the
comparison with M15 allows to infer that J1337-39 lies at the
extreme metal poor end of dwarf galaxies and its abundance has to
be lower than $ -2.17$.

Such a low abundance is consistent with the value obtained from
the optical spectroscopy if one assumes that the current
generation of stars has been enriched with respect to the RGB
population. We can also compare the CMD features with isochrones
at different metallicities to definitely constrain the possible
range of abundances and ages which can produce the observed colour
of the RGB, while being at the same time consistent with the
measured oxygen abundance.

\begin{figure}
\includegraphics[width=84mm]{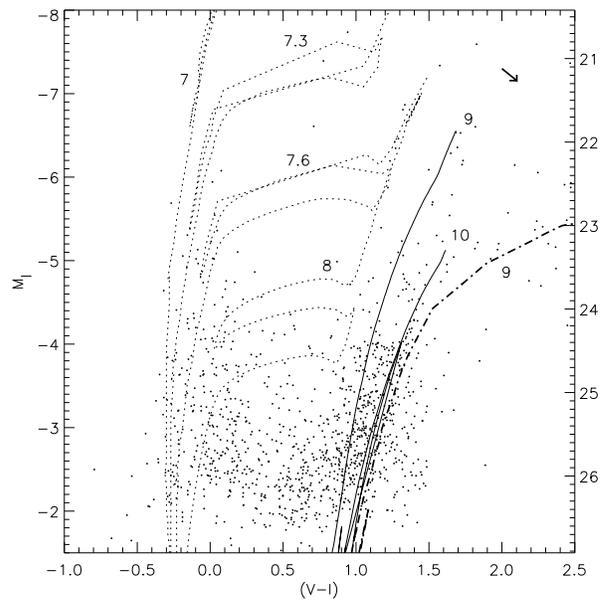}
%\vspace{63mm}
    \caption{HIPASS J1337--39: dereddened ($V-I$, $I$)  colour magnitude diagram overlaid with
    Padua evolutionary tracks at Z= 0.0004 ({\em solid lines}) and Z = 0.001 ({\em dotted lines}).
     The different isochrones correspond
    to $\log (Age) =$ 7 (10 Myr),  7.3 ($\sim$ 20 Myr) 7.6 ($\sim$ 40 Myr), 8 (150 Myr), 9 (1
    Gyr) and 10 (10 Gyr). The dashed-dotted line indicates an isochrone
    at Z= 0.004 of age 1 Gyr, to show what is the maximum
    metallicity allowed to fit the colour of the RGB. The arrow at the top-right corner indicates the reddening vector.}
    \label{39iso}
\end{figure}

Fig. \ref{39iso} displays the stellar tracks from the Padua group
(Bertelli et al 1994, Girardi et al. 2002). The set of isochrones
which best fit the observed colour of the RGB is at Z = 0.0004
(1/50 solar). This abundance corresponds to  a poor-metal scenario
with an intermediate-age (1 - 10 Gyr) dominant stellar population.
Fig. \ref{39iso} also shows an isochrone of age 1 Gyr at Z = 0.004
(dashed line) demonstrating that at higher metallicity  the bulk
of the RGB stars would have to be younger than 1 Gyr,
corresponding to a completely opposite scenario. This is the
maximum metallicity allowed at which isochrones can match the
position of the RGB in the diagram. However this possibility seems
highly unlikely, since it would be difficult to explain a
population of 1 Gyr old stars with a metallicity higher than the
current abundance of the ISM.

Therefore, by comparing the colour and the shape of the RGB with
the stellar evolutionary tracks we infer that the bulk of the red
stars observed is a metal-poor population which is $\lesssim$ 10
Gyr old. Thus, the age of this dwarf galaxy appears consistent with
formation in the early Universe rather than it being a recently formed system.
Further evidence, supporting the low metallicity --
intermediate-age population scenario will be discussed in the
following sections, such as the presence of a low metallicity
extended AGB branch.

\subsection{AGB stars: tracing the intermediate-age population}

The presence of low mass AGB stars in a galaxy is evidence for an
intermediate-age population and thus that SF activity has taken place
at look-back times of several Gyr. For this reason identifying AGB in
these gas-rich dwarfs would set another constraint on
their age and SF history. The distribution
of intermediate-age AGB stars in the CMD depends very strongly on
the metallicity: the higher the metallicity, the more extended
this feature is towards colours redder than the RGB. At
metallicities Z $\lesssim$ 0.001 the AGB extends almost vertically
above the RGB instead of bending towards redder colours (Cole et
al. 1999).

Candidate AGB stars in J1337--39 form a very narrow ``red tail''
elongated for roughly one magnitude above the tip with colours
similar to the RGB (1.2 $< V - I <$ 1.6). This would suggest the
presence of  a metal-poor, intermediate-age AGB population. The
isochrones at 10 Gyr and Z = 0.0004 (Fig.
\ref{39iso}) match the thin red tail of AGB stars and define an
approximate lower limit on the age of the galaxy. Using Z = 0.001
isochrones, the age of the AGB stars decreases to about 7 Gyr (not
shown in the figure to avoid confusion). Therefore
the AGB can be taken as a further argument in favour of the
low-metallicity, intermediate-age scenario previously inferred
from the RGB analysis.

We have also detected a number of very red stars at colours $V - I
\gtrsim 2.0$ and brighter than the TRGB (22 $< I <$ 24.5) whose
nature is unclear. We have looked at the CMD of the adjacent
fields to check if there are stars with similar photometric
properties ($V - I > 1.9$, 22 $< I <$ 24.5) and find 15 such stars
in the field, compared to 27 stars detected around J1337--39. The
probable contamination by foreground stars in the area is lower
than the number of stars observed (N$_{obs}$ $-$ N$_{foreground} =
12 \pm 6$), thus we can not exclude that they may belong to the
dwarf.

However, after having verified that they are point-source like in
appearance, we have checked their position in the galaxy and
compared their distribution to the other candidate AGB stars at
bluer colours forming the narrow red tail above the TRGB.
The redder AGB stars are mostly found at the border of the main
body of the galaxy, with a wider distribution than the bluer AGB
candidates.

The presence of these stars in HIPASS J1337--39 is a puzzle. If
they are AGB stars then their red colours are not easy to explain
and they imply the presence of a higher metallicity population.
Only isochrones of age between $\sim$ 1  and 3 Gyr at Z = 0.004,
in fact, can match their location on the CMD (Fig. \ref{39iso}),
but this would be in sharp contrast with the lower abundance of
the ISM that we have measured. If they belong to HIPASS J1337--39,
they would be younger than the bluer (and lower metallicity) AGB
candidate stars. If this were the case, how can  they more widely
spread throughout the frame than the older AGB stars, mostly
detected within or around the optical body of the galaxy?

Without further observations,
either spectroscopy or narrow band photometry using the $Ti-O$, or
$CN$ filters it is not possible to discriminate the nature of
these red stars in J1337--39 and to understand if they belong to the
dwarf galaxy rather than being foreground galactic stars.

To summarise, it appears from the red features in the CMD of
J1337--39, that the galaxy contains a dominant population which is
at least several Gyr old, with a lower limit on the age set by the
presence of candidate AGB stars. From the comparison with
isochrones that fit the position of the AGB stars, one obtains at
Z = 0.0004 an upper limit of $\sim$ 10 Gyr (or $\sim$ 7 Gyr if the
metallicity of the isochrones is Z = 0.001). The possible presence of a
population of very red AGB stars in HIPASS J1337--39 does not fit
very well with the scenario of a galaxy with an overall low metal abundance
and remains a puzzle.

\subsection{The recent SFH}

The evidence of recent star formation activity in HIPASS J1337--39
is clearly inferred from the population of young and blue stars at
colours $V - I \lesssim 0.5$ extending up to $I \simeq 22$ (Fig.
\ref{39iso}). This ``blue plume'' is populated by intermediate
mass stars in the main sequence (MS) phase and more evolved core
helium burning stars of the blue loop (BL) that form a redder
track parallel to the MS. The location of the different
evolutionary stages  is shown by
the isochrones at Z = 0.001 ({\em dotted line}) in %the left panel of
Fig. \ref{39iso}. We used a higher metallicity set of isochrones
for the recent stellar population to be consistent with the
results from the \hii region analysis.  There is evidence for
fairly continuous star formation activity between 20 and 100 Myr.
In fact the bright red supergiant stars with an absolute magnitude
$-6 > M_I > -7$ are fitted by the isochrone at $\log (Age) = 8$
and 7.6, while the positions of the brightest blue supergiants in
the diagram are consistent with isochrones around 20 ($\log (Age)
= 7.3$) and 40 Myr $\log (Age) = 7.6$.
%%%and the
%%%corresponding BL stars at the same age
%%The upper part of the CMD lacks bright stars, implying that there
%%have not been substantial SF episodes between 10 and 30 Myr ago.
The stars in the MS appear to be no older than 10 Myr.

\begin{figure}
\includegraphics[width=84mm]{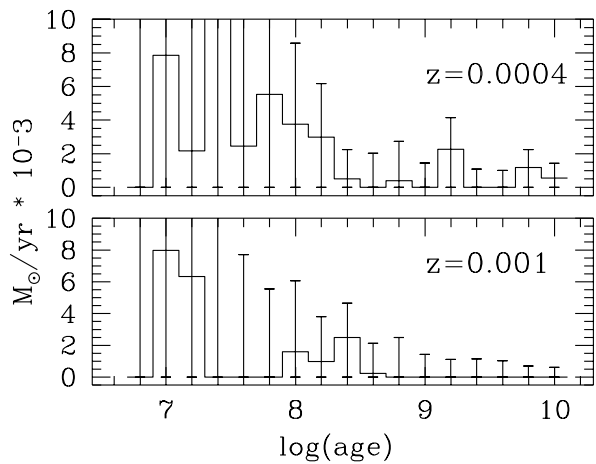}
%\vspace{84mm}
    \caption{The SFH of J1337--39 obtained with a combination of
    stellar isochrones with metallicities Z = 0.0004 and Z = 0.001. We
    find two main episodes of star formation: one between 6 and 10 Gyr
    ($\log (Age) = $9.8 - 10), and the other around 1.5 Gyr ($\log (Age)
     = $9.2). An increase in the SFR seems to occur in the last 200 Myr.}
    \label{bestSFH}
\end{figure}

\subsection{Modelling the SFH of HIPASS J1337--39}

We have used the {\scshape starfish} package of Harris \& Zaritsky
(2001) to investigate the SFH of the galaxy. First a linear
combination of synthetic CMDs is built by a set of stellar
isochrones, taking into account the scatter in the stars'
photometry as follows from artificial star tests (see \S 2.2 for
details), the interstellar extinction and the distance modulus of
the observational data. Then the  code compares the synthetic and
the observed CMDs. It determines the best-fit SFH performing
$\chi^2$ minimization of the differences in the number of stars
between the model and the data in the different regions of the
diagram.

We used the set of isochrones provided by the Padua group
with different values of the metallicity. %to build synthetic CMDs of J1337--39.
We considered the entire range of available ages (from $10^7$ yr
to $10^{10.2}$ yr) with a step of $\log (Age) = 0.2$. We adopted a
Salpeter initial mass function(IMF), a binary fraction of 0.5, and
we included Galactic foreground extinction (but not any internal
extinction in the dwarfs).

For HIPASS J1337-39 the distance modulus was set to $(m - M)_0 =
28.42$. The metallicity is constrained between 1/50th and 1/30th
solar from the analysis of the CMD and the oxygen abundance.
Therefore we have created synthetic CMDs with various combination
of isochrones at Z = 0.0004, Z = 0.001 and also  at Z = 0.004 to
test the effects of a higher metallicity component. In Table 3 we
show the fits we obtained for different metal abundances.

The resulting SFH (Fig. \ref{bestSFH}) which gave the lowest
reduced $\chi^2$ value  and the best agreement between the
observed and synthetic CMDs corresponds to the combination of the
isochrones at Z = 0.0004 and Z = 0.001. From the figure there is
evidence of an extended period of SF around 6 - 10 Gyr ($\log(Age)
=$ 9.8, 10) for Z = 0.0004. The galaxy then appears to go through
a rather quiescent period until about 1.5 Gyr ago. For ages around
or less than 1 Gyr the SFR seems to be rather low except for the
remarkable enhancement in the last 200 - 250 Myr. The oldest stars
have metallicities Z = 0.0004, while populations with higher metal
abundance appear only in the last Gyr.

\begin{table}
\caption{The list of the different combination of metallicities
inspected with {\scshape starfish} and the corresponding reduced
$\chi^2$ parameter. For HIPASS J1337-39 we find that the best-fit
SFH is given by a model with isochrones at Z = 0.0004 and 0.001.}
\begin{center}
\begin{tabular}{|ccc|}
  \hline \hline
  % after \\: \hline or \cline{col1-col2} \cline{col3-col4} ...
  Z &  & $\chi^2$ \\
  \hline \hline
  0.0004 &  & 1.25\\
  0.001 &  & 1.33 \\
  0.004 &   & 1.34\\
  0.0004+0.001 & & 1.08  \\
  0.001+0.004 &  & 1.28 \\
  0.0004+0.001+0.004 & & 1.13 \\
  \hline \hline
\end{tabular}
\end{center}
\end{table}

The error bars are determined by the code identifying the 68\%
(1$\sigma$) confidence interval on each amplitude. The number of
stars in the diagram is low, especially the young blue ones, which
makes the statistical analysis and modelling extremely difficult
in this part of the diagram. For this reason the uncertainties in
the SFR estimation are rather large at recent epochs.

\begin{figure*}
\begin{minipage}{126mm}
\includegraphics[width=63mm]{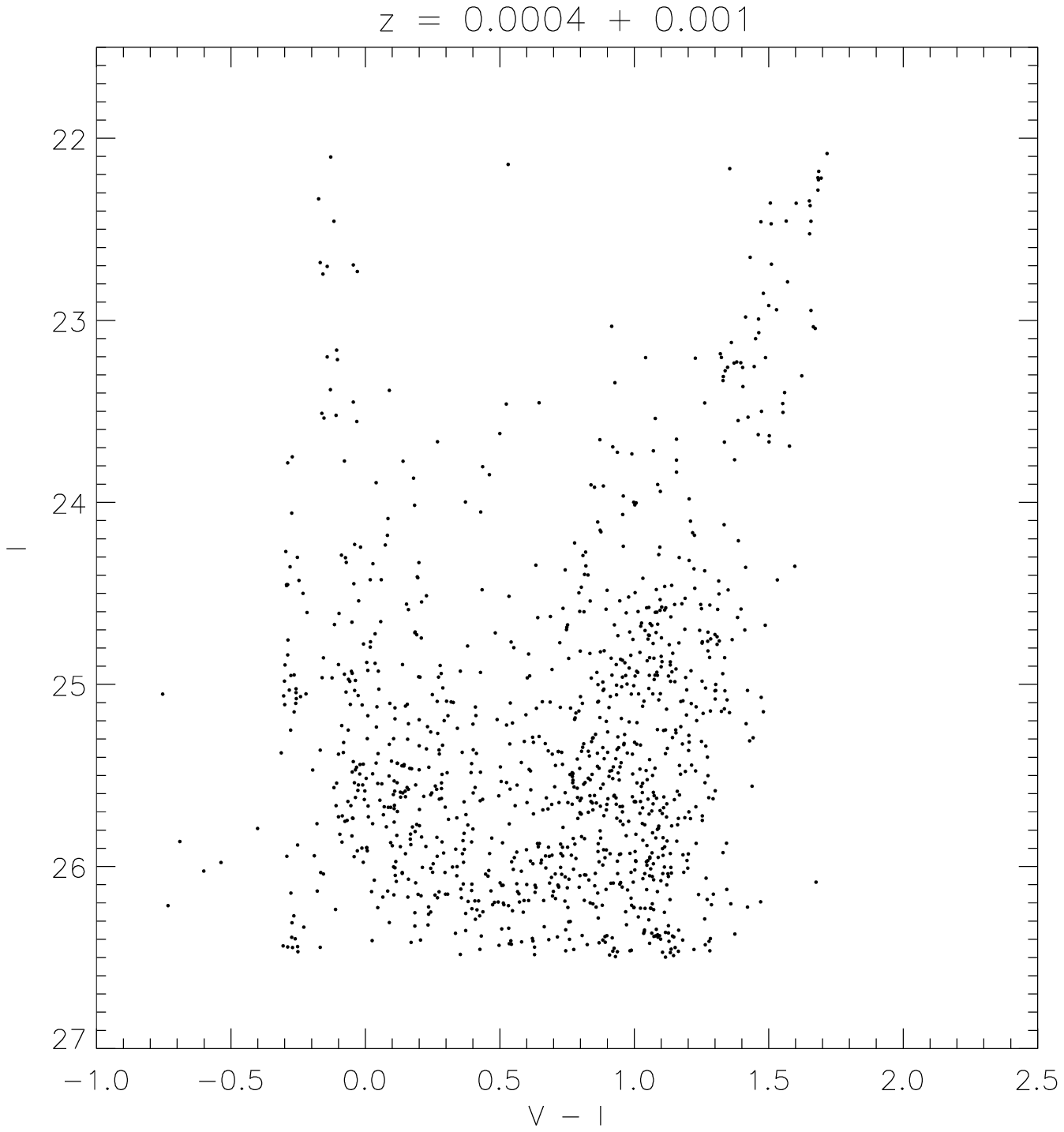}
\includegraphics[width=63mm]{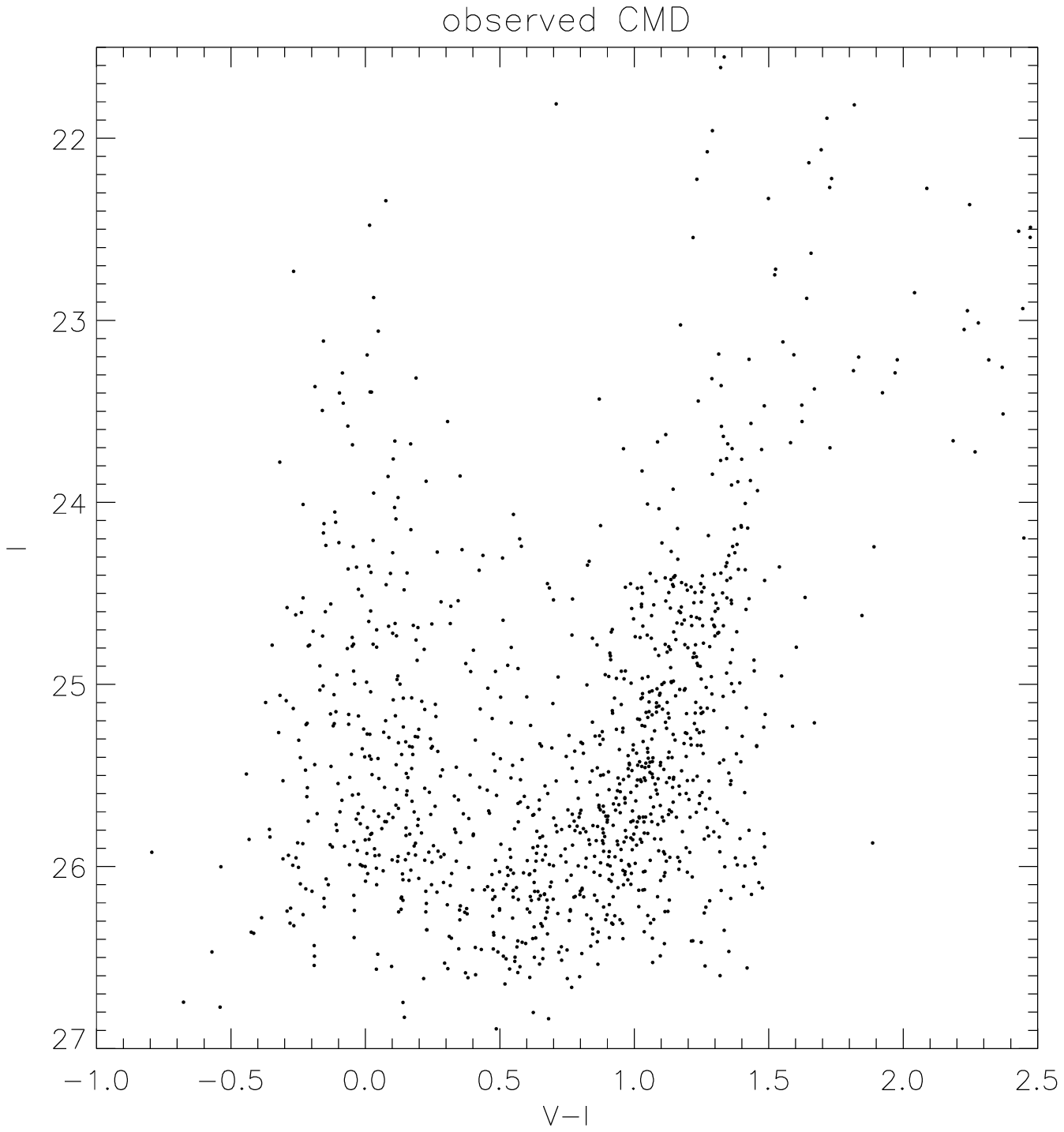}
%\vspace{63mm}
    \caption{The comparison between the model {\em (left)} and the observed CMD of HIPASS J1337--39 {\em (right)}.}
    \label{SFISHoutput}
\end{minipage}
\end{figure*}

The corresponding model CMD is displayed in Fig. \ref{SFISHoutput}
and compared to the observed data. The two diagrams look fairly
similar. The average colour of the simulated RGB is consistent
with the observed one. The main differences appear in the
modelling of the red stars above the tip of the RGB.  A few
intermediate-age AGB stars are found in the two oldest age bins
($\log(Age) =$ 9.8, 10)
but they are not distributed as a narrow plume just above the TRGB
as in the data. Moreover too many red bright stars
with $22 < I < 24$ and colours $1.2 < V - I < 1.7$ appear in the
age bin corresponding  to $\log(Age) =$ 9.2, indicating that the
SFR in that bin has been overestimated by the code.

A model CMD built using only the highest allowed metallicity, Z =
0.004 fails to reproduce the observed diagram and the
corresponding fit has a higher $\chi^2$ value.

To conclude, the simulation of the SFH of HIPASS J1337--39 points toward
a scenario with a low metallicity population with an age as old as 10
Gyr, in agreement with our analysis of the CMD.

\subsection{Summary and discussion}

Our results show that HIPASS J1337--39 is not a young galaxy.  To
explain the colour of the RGB with a population of young stars
would require a metallicity of around 1/5 solar (Z = 0.004), which
would be inconsistent with the metal abundance measured in the
ISM.  On the other hand, comparison of the CMD with theoretical
low-metallicity isochrones indicates that the RGB may contain
stars as old as 10 Gyr.  The detection of AGB stars gives further
evidence that the galaxy may harbour an intermediate-age stellar
population.

The synthetic CMD is able to reproduce the main features of the
observed one.  Again, this implies that the first episode of SF
occurred $10^{10\pm 0.1}$ years ago.  We find an average SFR over
10 Gyr of $\sim 2 \times 10^{-3}$ \msun yr$^{-1}$, and the SFH of
J1337--39 appears to be characterised by three main episodes of SF
(between 6 and 10 Gyr, around 2.5 Gyr and in the last 250 Myr).

Nevertheless, to firmly prove the existence of an old stellar
component one should unearth a population of horizontal branch
(HB) stars, which would require much deeper observations with the
HST.

If our interpretation is correct and this galaxy harbours a
stellar population older than a few Gyr, the low oxygen abundance
that we have measured is remarkable as it indicates
that the galaxy has undergone little chemical enrichment during
its evolution.

We have also analysed the neutral gas content of J1337--39 to
investigate its distribution and its relation with the SF
activity. The galaxy has an overall \hi column density which is
above $10^{20}$ \cmdue until 2 kpc from the centre. The \hi peak
density (few times 10$^{21}$ \cmdue) coincides with the area where
the \hii regions are found. The \hi distribution around J1337--39
appears regular overall, apart from an unexplained ``tail'' to the north
west.

There do not seem to be any objects within the 30\arcmin $\;$
field of view of the ATCA telescope to explain the slightly
disturbed \hi morphology as the effect of a tidal interaction with
a close companion. Moreover, at the distance of 4.8 Mpc that we
derived from the tip of the RGB, J1337--39 seems to be located on
the margins of the Centaurus A group. The galaxy is about
10\arcdeg ($\sim$ 850 kpc) from M83, the closest large galaxy in
the Centaurus A group. Among the other massive galaxies of the
group, NGC 5102 is at a radial distance d = 3.4 Mpc, while NGC
5253 is at $d$ = 3.9 Mpc (Karachentsev et al. 2002) at an angular
separation of about 8 degrees -- both of these are well over 1 Mpc
from J1337--39.

The shape of the \hi distribution could be due to the interaction
with a hot intracluster medium. A galaxy moving with a velocity
$v_G$ through a hot diffused gas is subjected to a ram pressure
given by $p_{ram} = \rho_{ICM} v^2_G$, where $\rho_{ICM}$ is the
density of the medium. However this seems unlikely as the hot gas
density should be lower at the edge of the group and J1337--39 is
one of the most distant dwarfs from M83.

J1337-39 is at only 4 degrees in projection from NGC 5128, which
is the nearest active galactic nucleus. NGC 5128 shows both large
radio lobes covering at 1410 MHz an area of $8^{\circ} \times
4^{\circ}$ in the north south direction (Cooper et al. 1965;
Junkes et al. 1993), and extended X-ray emission ($7^{\circ}
\times 1.5^{\circ}$)  along the direction north east - south west,
rotated about 50 degrees  from the axis of the radio lobes (Arp
1994). The region  of X-ray emission around Cen A corresponds to
roughly 400 kpc, and indicates the existence of a diffuse hot gas
($\sim 10^7$ K) (Arp 1994). From the Rosat map, this emission
would include the area of the sky where J1337-39 is found. However
the relative distance between the two galaxies is about 1.4 Mpc,
assuming that $d_{CenA} = 3.66$ Mpc and $d_{39} = 4.8$ Mpc.
Without any indication on the effective direction of the X-ray jet
with respect to J1337-39, and given the large relative distance
between the two galxies is difficult to argue that the \hi
morphology of this dwarf is an indication of ram pressure
stripping interaction with NGC 5128.

Alternatively, the \hi distribution may be the result of a recent
accretion of gas which would be consistent with the irregularities
in the velocity field in the \hi tail. The possible accretion of
an \hi cloud may be related to the apparent increase in the SF
activity around 100  Myr ago that was inferred from the analysis
of the model and observed CMDs. Higher resolution \hi data are
needed to investigate the structure of the gas distribution in
J1337--39 in more detail.

\section{HIDEEP J1337--3320}

HIDEEP J1337--3320, which was found in a deeper \hi survey than
the other dwarfs, has the lowest \hi mass, $M_{HI} = 5 \times
10^6$,
 and column density (see Table \ref{tabprop}). Its gas fraction is also lower than the others (\ml
$=$ 1.4 in solar units).  Optically, we find a small and rather
compact object %%n LSB counterpart ($\mu^B_0 = 23.70 \pm 0.04$ mag
%%\arcsecdue)
with no evidence for ongoing star formation.  We are
therefore unable to independently constrain the metallicity by
measuring the oxygen abundance, and thus are faced with
difficulties in the analysis of the CMD due to the age-metallicity
degeneracy.

\subsection{The neutral gas content}

\begin{figure*}
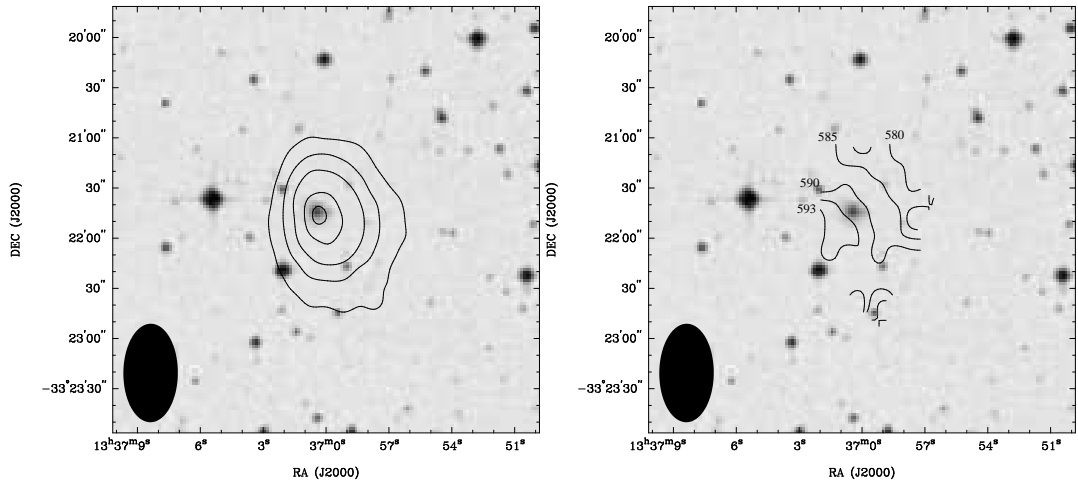

\begin{minipage}{126mm}
\includegraphics[width=63mm,angle=-90]{Figure11a.eps}
\includegraphics[width=63mm,angle=-90]{Figure11b.eps}
%\vspace{63mm}
    \caption{{\em Left:} \hi density contours overlaid on the DSS field including the galaxy. The contour levels
are: 0.5, 1.0, 1.5, 2.0, 2.25 $\times 10^{20}$ \cmdue. {\em
Right:} The velocity field displayed in the range v = 580 \kms and
$v$ = 593 \kms . The beam size is shown at the bottom left corner
of the image. }
    \label{33HI}
\end{minipage}
\end{figure*}

HIDEEP J1337--3320 has a very smooth \hi distribution that is
centred on the stellar component of the galaxy (Fig. \ref{33HI}),
although the presence of local clumps and perturbations could be
smeared out by the large beam size, which only marginally resolves
the galaxy. The neutral gas extends beyond the optical counterpart
out to 45\arcsec, $\sim$ 1 kpc at an assumed distance of 4.4 Mpc
(see below), and can be traced out to a column density $N_{HI} = 5
\times 10^{19}$ \cmdue. The gas density around the stellar
distribution is almost constant, but yet it is low, hovering
around $N_{HI} = 2 \times 10^{20}$ \cmdue with a central peak at
$N_{HI} = 2.3 \times 10^{20}$ \cmdue. In general dIrr star forming
galaxies in the LG show more asymmetric \hi distributions with
holes and arcs as in IC 10 (Wilcots \& Miller 1998) or SagDIG
(Young \& Lo 1997) that may be created by supernovae or stellar
winds. The smooth distribution of the gas may give hints of a lack
of star formation in the last $\sim$ 100 Myr, however, one has to
keep in mind that the low resolution of the observations may have
smeared out irregularities in the \hi structure. There is some
evidence of a velocity gradient both from the velocity contours
(Fig. \ref{33HI}, right panel) and from the individual channel
images (not shown here).

The \hi profile is centred at 592.1 $\pm$ 0.8 \kms , with a 20\%
velocity width $\Delta v_{20} = 38$ \kms and a velocity dispersion
$\sigma_v \simeq 8$ \kms . We derive an \hi mass of $M_{HI} = 5.1
\times 10^6$ $M_{\odot}$, at a TRGB distance of 4.4 Mpc (see
below). The dynamical mass, without any inclination angle
correction, is $M_{dyn} \simeq 8 \times 10^7$ \msun, giving
$M_{HI}/M_{dyn} \simeq 0.06$.

\begin{figure}
\includegraphics[width=84mm]{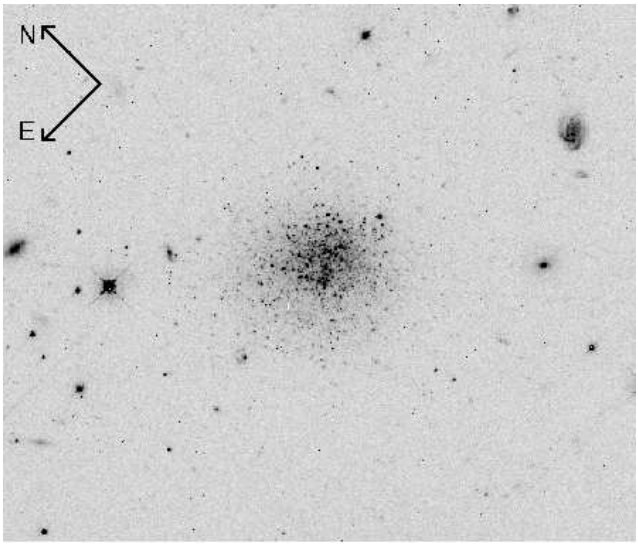}
%\vspace{84mm}
    \caption{The WFPC2 image of HIDEEP J1337--3320 taken with the filter F555W. The size of the frame is
roughly 80\arcsec. }
    \label{33image}
\end{figure}

\subsection{Optical properties}

The optical image of HIDEEP J1337-­3320 (Fig. \ref{33image}) shows a smaller
and more compact object than J1337-­39. It does not seem to have signs
of ongoing star
formation activity and the WIYN snapshots did not show evidence
for \hii regions. The stellar distribution
looks very smooth and regular, and its surface-brightness
distribution can be fitted by an exponential profile.

Its optical morphology and the presence of gas (\ml = 1.4)
resemble the transition type dwarfs of the Local Group such as LGS
3, Phoenix, and DDO 210 (although they have much lower \ml ratios)
and the more gas-rich dwarfs with intermediate properties between
dSphs and dIrrs found by Skillmann, C\^ot\'e \& Miller (2003,
hereafter SCM03) in the Sculptor group.

\subsection{The morphology of the CMD}

In the left panel of Fig. \ref{33CMD} the ($I$, $V - I$)
morphology of the CMD is shown. The main features at red colours
are a narrow RGB, a few candidate AGB stars, some of them with
colour ($V - I$) $> 2$. The blue part of the diagram, on the other
hand, is less populated, especially if compared to J1337-­39, and
from the comparison with theoretical isochrones (see section 4.6)
it seems to contain mostly blue loop stars, suggesting no SF
activity within the last 10 - 30 Myrs.
%%%In the right panel we
%%%display the CMD of the fields (WF2, WF4 chips) adjacent to the
%%%galaxy which can be used as a comparison to check the galactic
%%%foreground contamination.

\subsection{The RGB, and the distance to HIDEEP J1337-­3320}

The RGB is the most prominent feature of the CMD.  It is slightly
redder than that of J1337-­39 and is located between $V - I$ = 1
and $V - I$ = 1.6.  As with J1337--39, the distance to the dwarf
can be determined from the $I$ magnitude of the tip of the RGB.
This is found at $I_0 = 24.15 \pm 0.10$ (Fig. \ref{33lum})
assuming a galactic reddening $E_{B-V} = 0.049 \pm 0.008$ and $A_I
= 0.10 \pm 0.01$. The de­reddened distance modulus is $(m - M)_0 =
28.20 \pm 0.10$ and the resulting distance to HIDEEP J1337-­3320
is $D = 4.4\pm 0.2$ Mpc.

At an angular separation of about 3.5\arcdeg $\,$ from M 83
($\lesssim$ 300 kpc at 4.5 Mpc) and a comparable radial distance
($d_{M83} = 4.5$ Mpc) HIDEEP J1337-­3320 seems to be located in a
position within the group that is much closer to the massive
spiral than J1337-­39.

\begin{figure}
%\begin{figure*}
%\begin{minipage}{126mm}
\includegraphics[width=87mm]{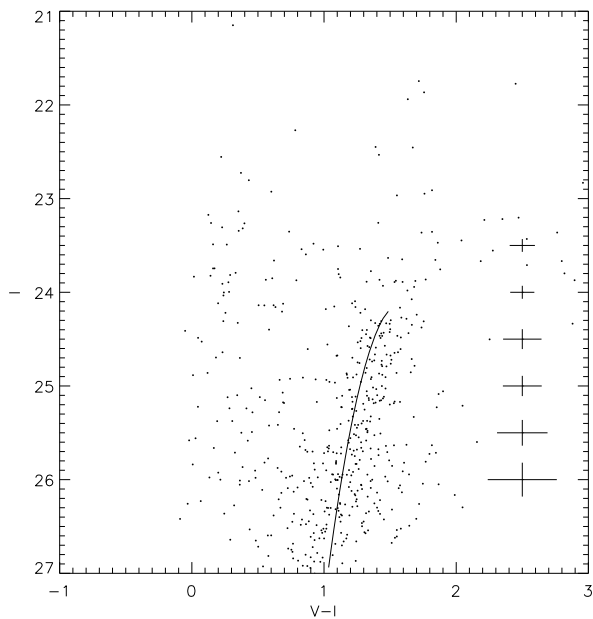}
%\includegraphics[width=63mm]{Figure13b.eps}
%\vspace{63mm}
    \caption{HIDEEP J1337-3320: $(V - I)$, $I$ color magnitude diagram.
 The solid line correspond to the RGB of the galactic globular cluster  M 15 and its location
  shows that the metallicity
 of J1337--3320 is on average slightly higher than M 15 (\fe = $-2.1$).}
 %%%{\em Right:} The CMD of the fields in the
%%%  WF2 and WF4 chips
%%%    where the galaxy was not included. The diagram has been
%%%    constructed  by selecting the halves of the chips which were not attached to the WF3.}
    \label{33CMD}
%\end{minipage}
\end{figure}
%\end{figure*}

In Fig. \ref{33CMD} we have overlayed on the CMD of J1337-­3320 the
RGB of the galactic globular cluster M 15, suitably scaled to the
distance of 4.4 Mpc. The colour of the RGB of M 15, with a
metallicity \fe = $-2.17$, is slightly bluer than the mean
distribution of the observed RGB. Therefore, if
we assume that the galaxy
has a population of stars which is at a few Gyr old, we can obtain a
lower limit on the average abundance of J1337-­3320. Using the
method of Lee et al. (1993) to estimate the metallicity from the
colour of the RGB, we find \fe = $-1.9 \pm 0.6$, slightly
higher than HIPASS J1337-­39.

When the CMD is compared with Padua stellar evolutionary tracks,
we find that the Z = 0.001 set provides the best
fit to the RGB. In this scenario, the population of the RGB
appears to be as old as 10 Gyr. However, a different choice of the
metallicity changes the upper limit on the age one can derive from
the RGB. If we use the Z = 0.004 (1/5 solar) isochrones, as
shown in Fig. \ref{33iso}, the oldest track which matches the RGB
is at about 4 Gyr. Tracks with ages around and older than 1 Gyr
can also fit the position of the very red stars (possible AGB
candidates) at $V - I >$ 2.
In this scenario, the blue part of the RGB (i.e. stars with $1 < V-I < 1.6$
and $I > 24.2$) would be very young, with an age between 250 Myr and 1
Gyr, as one can see from the right panel of Fig. \ref{33iso}. The
age of the onset of the helium flash is between 0.9 and 1.5 Gyr,
therefore the stars with
Z = 0.004 and ages
 $\lesssim$ 1 Gyr would be ascending
the AGB for the first time, rather than being in the red giant
phase. However AGB stars do not show a clear-cut cutoff in
luminosity as RGB stars do, thus, in this scenario, a sharp
decrease in the number of stars at $I$ $\sim$ 24.2, i.e. the tip
of the RGB, would probably not be detectable in the diagram - as
it clearly is.

\begin{figure}
\includegraphics[width=74mm]{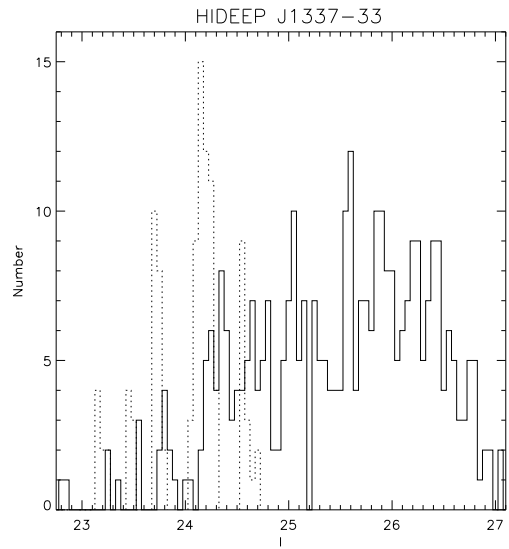}
%\vspace{84mm}
    \caption{The $I$ luminosity functions of HIDEEP J1337--3320. The dotted line represents the
    output of the edge detection filter used to detect the tip. The
    peak indicates the position of the tip which has been found at $I_0$ = 24.15 $\pm$ 0.1}
    \label{33lum}
\end{figure}

The shape of the red branch and the presence of a clear cut-off in the
luminosity function around $I \sim 24.2$ can be taken as
evidence of the existence of an intermediate-age population of red
giants. A similar argument has been used in other systems (such as
Leo A; Tolstoy et al. 1998) when more reliable age indicator have not been
detected in the CMD.

\begin{figure*}
\begin{minipage}{126mm}
\includegraphics[width=63mm]{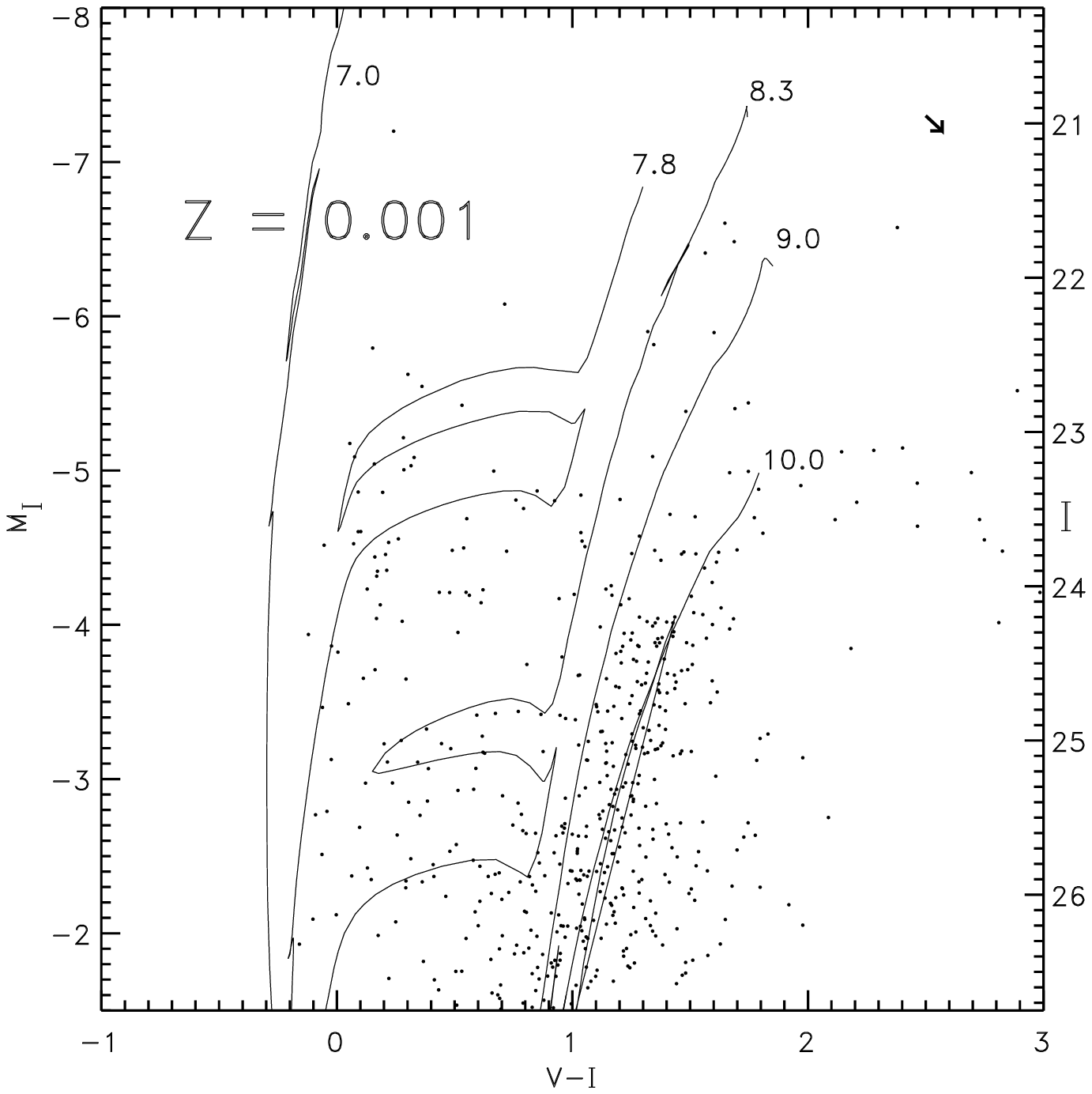}
\includegraphics[width=63mm]{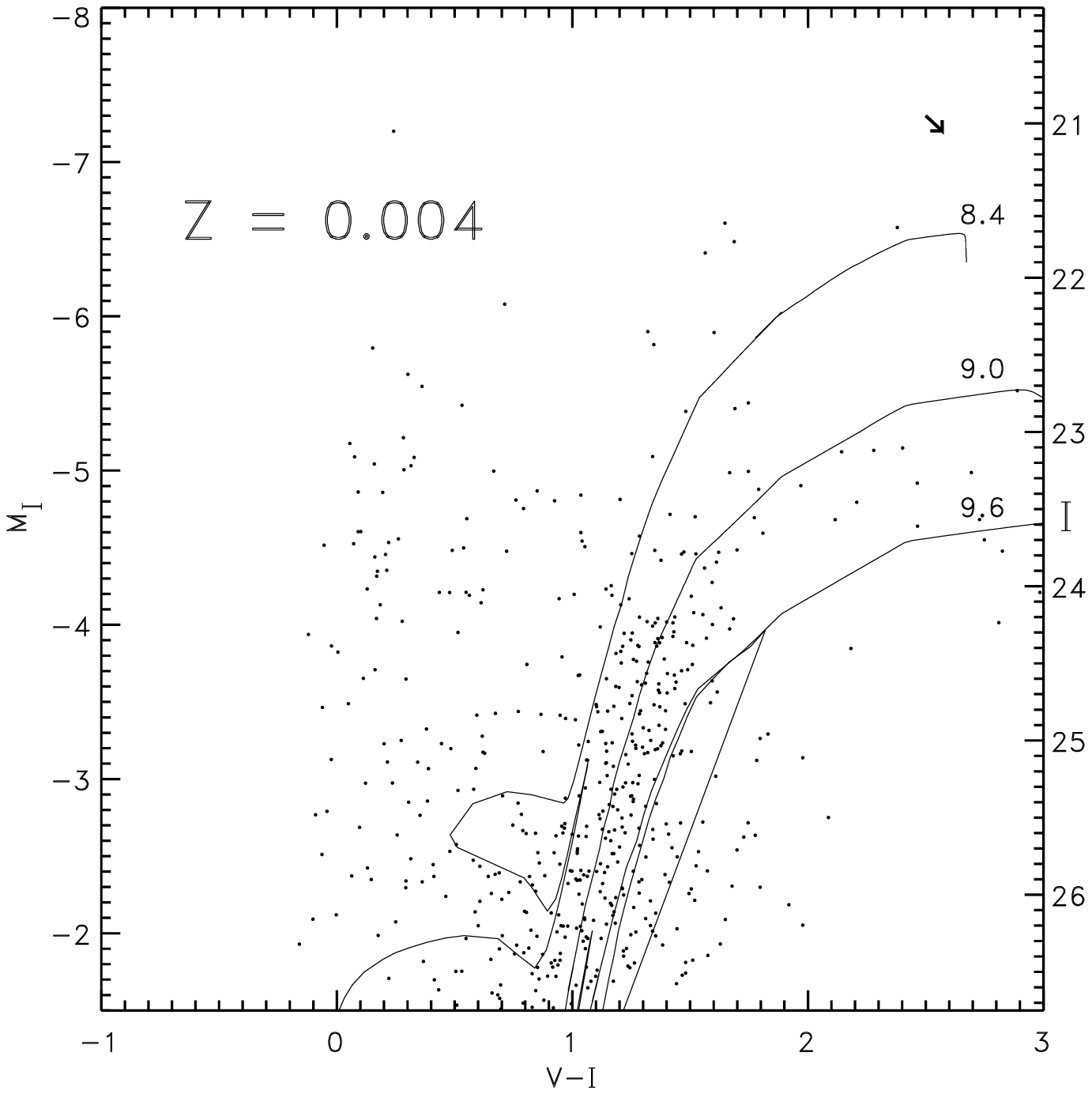}
%\vspace{63mm}
    \caption{HIDEEP J1337-­3320: dereddened $(V - I)$, $I$ color magnitude diagrams
overlaid with isochrones at Z = 0.001 {\em (left)} and Z = 0.004
{\em (right)}. According to the metallicity used, the limit one
can set on the age of the red giant population changes from 10 to
4 Gyr. At Z = 0.004, the blue part of the feature that we define
as the RGB (i.e. stars with colours between 1 and 1.6 and
luminosity fainter than $I = 24.2$) would be younger, with an age
between 250 Myr and 1 Gyr, therefore it would be mostly populated
by AGB stars rather than red giants. The age of the younger blue
stellar population, which are mainly core helium burning stars is
 shown in the left panel. Blue and very young main sequence stars
(age less than 10 Myr) seems to be lacking. The arrows at the
top-right corners indicate the reddening vector.}
    \label{33iso}
\end{minipage}
\end{figure*}

\subsection{The AGB}

A small number of objects, potentially AGB stars, have been found
above the TRGB (Fig. \ref{33CMD}), some of them (24 $\pm$ 5) with
the very red colours ($V - I > 1.7$) we found in J1337-­39.  If
these stars belong to J1337--3320, then they should have a
metallicity greater then 1/20 solar (Z=0.001), at which, according
to synthetic models, the AGB branch starts bending horizontally
towards the reddest part of the diagram (Fig. \ref{33iso}),
revealing the presence of a more metal-rich population in the
galaxy.  The local contamination around $V - I >$ 1.7, appears to
be low. Indeed in the adjacent fields of the camera where the
galaxy was not included (WF2 and WF4), we have detected 11 $\pm$ 3
stars  with $V - I
> 1.7$ and $22 < I < 25$.

Most of the candidate AGB stars are found within or around the
main optical body of the dwarf. Some of the reddest stars,
however, are clearly outside the bulk of the stellar distribution.
They are likely to be foreground stars, but we cannot exclude that
they belong to the dwarf since  there are examples in the LG
galaxies, such as IC 1613 (Albert, Demers, \& Kunkel 2000), DDO
210 and Pegasus (Battinelli \& Demers 2000), Leo I (Demers \&
Battinelli 2002) where Carbon stars extend even further out than
the observed \hi content.

It is possible that there is a population of
old AGB stars with colours similar to that of the TRGB whose
position on the CMD is fitted by the same low metallicity isochrones (Z =
0.001) as the RGB.  However, only isochrones at Z = 0.004 (1/5 solar)
can fully match their position on the diagram, thus it could be that
we have found a population of very red and bright AGB stars formed
at a later epoch with a higher metallicity.  The
very red AGB population may give a hint of the chemical evolution
of the galaxy through its different episodes of star formation.
The isochrones at Z = 0.004 also match the position of the
fainter and bluer AGB stars, thus we cannot be certain that there is only
a single population of relatively young and metal-rich AGB stars.
%%With no detected \hii regions we cannot measure the metallicity of
%%the gas.
Therefore spectra or narrow-band photometry are needed to measure
their metallicity in order to improve our understanding of the
reddest stars in J1337-­3320.

\subsection{The younger stellar populations.}

The blue plume of HIDEEP J1337-­3320 (Fig. \ref{33CMD}) is much
less populated than that of J1337-­39. There are few bright stars
up to $I \sim 22.5$, which is likely to be the bluest extent of
the helium burning phase from comparison with the theoretical
isochrones (Fig. \ref{33iso}). A well defined main sequence though
is lacking which  implies a drop of the recent SF activity.

Isochrones at Z = 0.001 fitting the magnitudes and colour of the
brightest blue loop stars indicate that the age of the most recent
stars in this galaxy is around 60 -- 100 Myr, when the last episode
of significant SF activity probably occurred.  %%%The most likely
%%%scenario is that the ``fading'' blue plume is due to a recent decrease in
%%%the SFR.

\subsection{SFH of HIDEEP J1337-­3320}

\begin{figure}
\includegraphics[width=84mm]{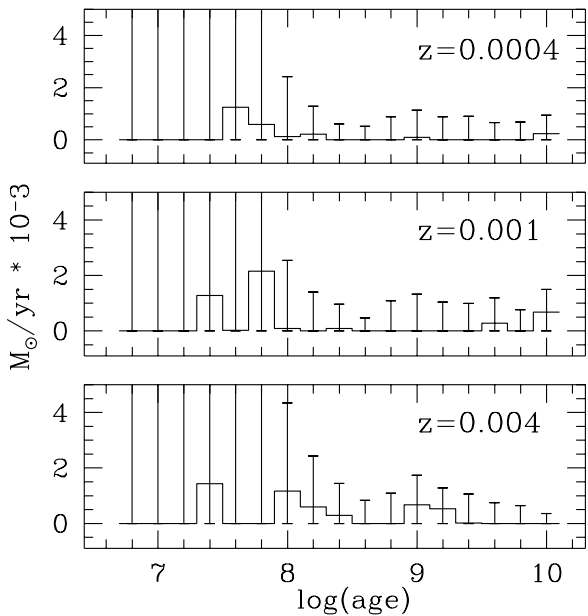}
%\vspace{84mm}
    \caption{The best fit SFH of J1337-­3320.}
    \label{33SFH}
\end{figure}

\begin{figure*}
\begin{minipage}{126mm}
\includegraphics[width=63mm]{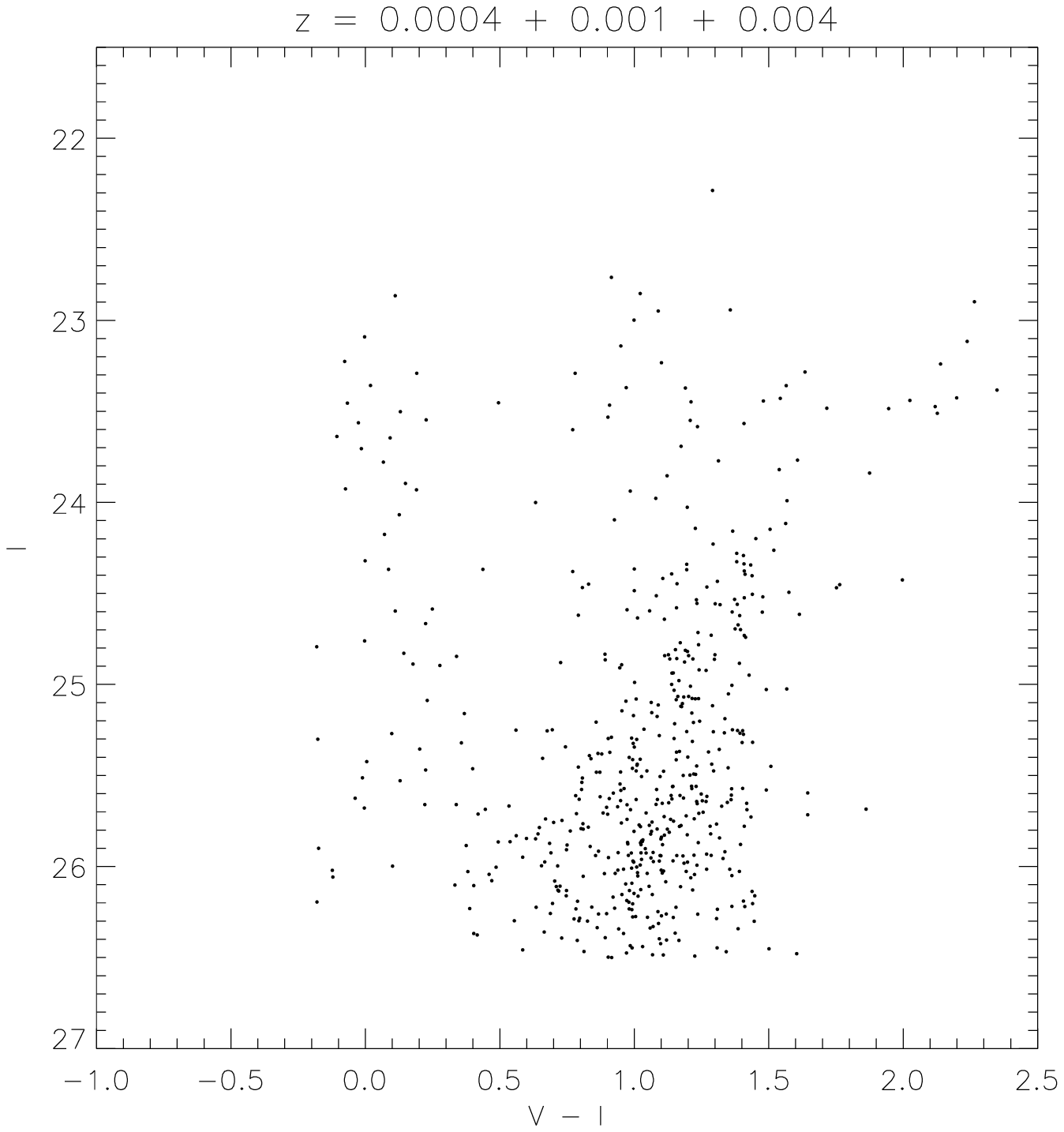}
\includegraphics[width=63mm]{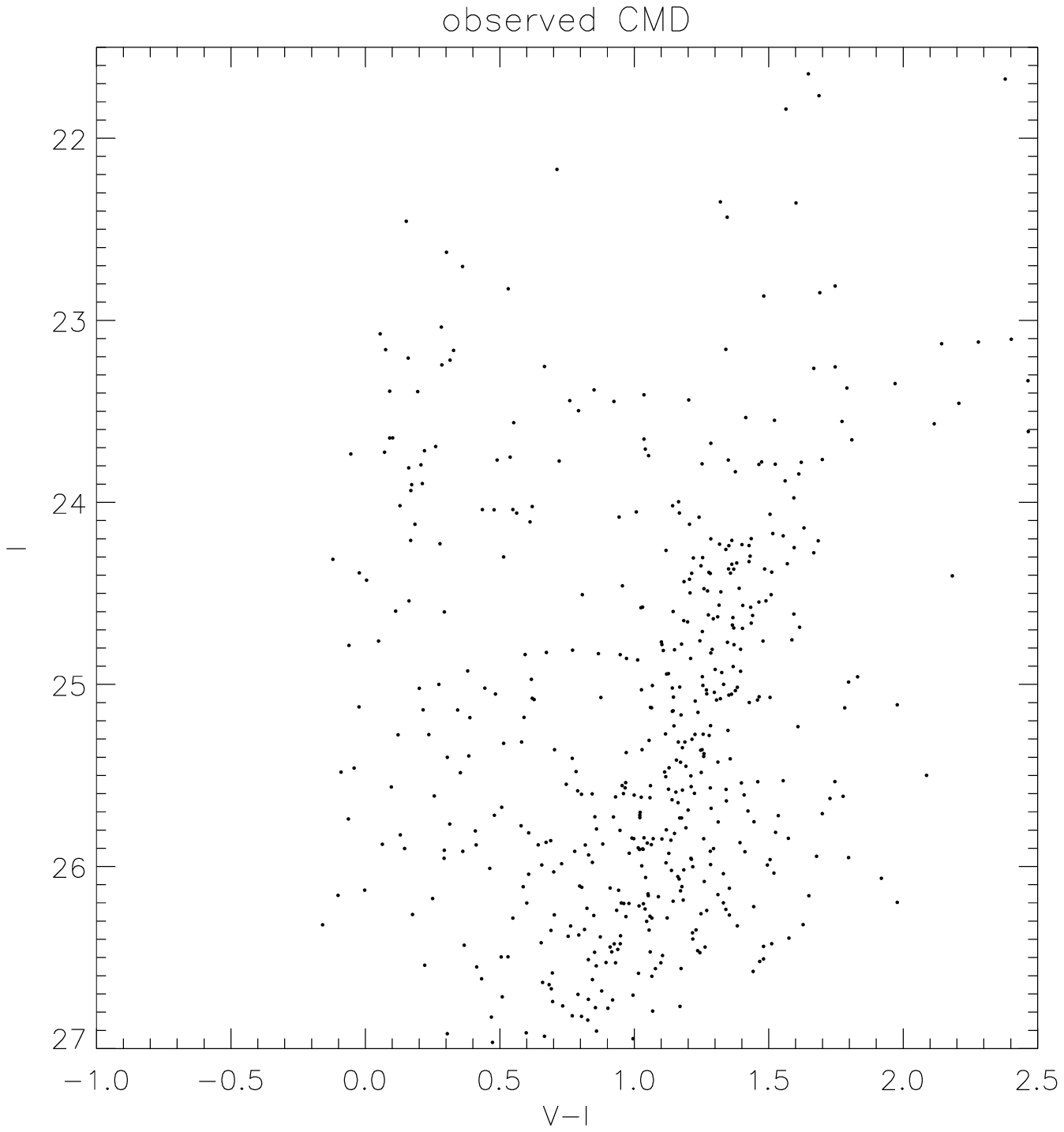}
%\vspace{63mm}
    \caption{The best fit model CMD ({\em left}) compared to the observed one ({\em right}). We find a good agreement between the two
diagrams although there is an overestimate of the old SFR which
produces a more populated RGB at faint magnitudes.}
    \label{33modelCMD}
\end{minipage}
\end{figure*}

We have used {\scshape starfish} to model the SFH of J1337-­3320.
The different combination of metallicities inspected and the
corresponding reduced $\chi^2$ values are shown in Table 4. The
best-fit SFH and the model CMD are displayed in Fig. \ref{33SFH}
and in the left panel of Fig. \ref{33modelCMD} respectively. The
observed CMD is also shown in the right panel of Fig.
\ref{33modelCMD} for comparison with the synthetic diagram. The
best fit is the result of the combination of three sets of
isochrones, Z = 0.0004, Z = 0.001, and Z = 0.004.

\begin{table}
\caption{The list of the different combination of metallicities
inspected with {\scshape starfish} and the corresponding reduced
$\chi^2$ parameter. For HIDEEP J1337-3320 we find that the
best-fit SFH is given by a model with  isochrones at Z = 0.0004,
0.001 and 0.004.}
\begin{center}
\begin{tabular}{|ccc|}
  \hline \hline
  % after \\: \hline or \cline{col1-col2} \cline{col3-col4} ...
  Z &  & $\chi^2$ \\
  \hline \hline
  0.0004 &  & 1.47\\
  0.001 &  &  1.22\\
  0.004 &   & 1.36\\
  0.0004+0.001 & & 1.26  \\
  0.001+0.004 &  & 1.23 \\
  0.0004+0.001+0.004 & & 1.15 \\
  \hline \hline
\end{tabular}
\end{center}
\end{table}

The evolution
of J1337-­3320 seems to be dominated by an early SF event at
$10^{+2.5}_{-2.0}$ Gyr where the majority of the red giant stars
have been formed. After this episode the galaxy appears to have
experienced episodes of SF occurring around 4 Gyr  and 1 -- 1.5 Gyr
followed by periods of apparent
inactivity. SF seems to be triggered again around 250 Myr
with an increased SFR until around 25 Myr, peaking between 60 and 100 Myr.
At ages younger than 100 Myr the errors in the SFR become very large,
affecting the results of the model for the youngest stars.

The overall SFR appears to be very low, and it never exceeds $2
\times 10^{-3}$ \msun yr$^{-1}$. There is a degeneracy at 10 Gyr
between the isochrones at Z = 0.0004 and Z = 0.001. Synthetic CMDs
with Z = 0.0004 and Z = 0.001 are photometrically too similar for
the method to be able to distinguish between them with the current
data. We observe a consistent age-metallicity relation in the SFH.
The oldest stars have Z = 0.0004/0.001 while there is no
significant population with Z = 0.004 before log (Age) = 9.2 which
suggests a gradual metal enrichment due to the previous SF
episodes.

The model CMD reasonably matches the main features of the observed
one. The main difference found is in the width of the RGB, which
is too large at faint magnitudes  in the model, probably due to an
overestimate of the SFR in the 10 Gyr age bin. From our
simulation, the majority of the red giant stars appear in the 10
Gyr interval. On the other hand, the recent SFR seems to be
slightly underestimated, with a small difference in the number of
young stars between the model and the data. The synthetic CMD also
show a red tail of stars above the TRGB extending at colours $(V -
I) > 1.5$, which appears at $\log$($Age$) = 9, 9.2 with a higher
metallicity (Z = 0.004).

To conclude, the evolution of J1337-­3320 seems to be characterised
by periods of low SF activity followed by quiescence
(the so called ``gasping'' evolution). The average SFR is always
below $2 \times 10^{-3}$ \msun yr$^{-1}$. At more recent epochs,
we find an increase in the SF activity within the last 200 Myr,
with a peak around 60 - 100 Myr, but there is no evidence of young
stars with ages less than 10 Myr.

\subsection{Summary and discussion}

\begin{figure*}
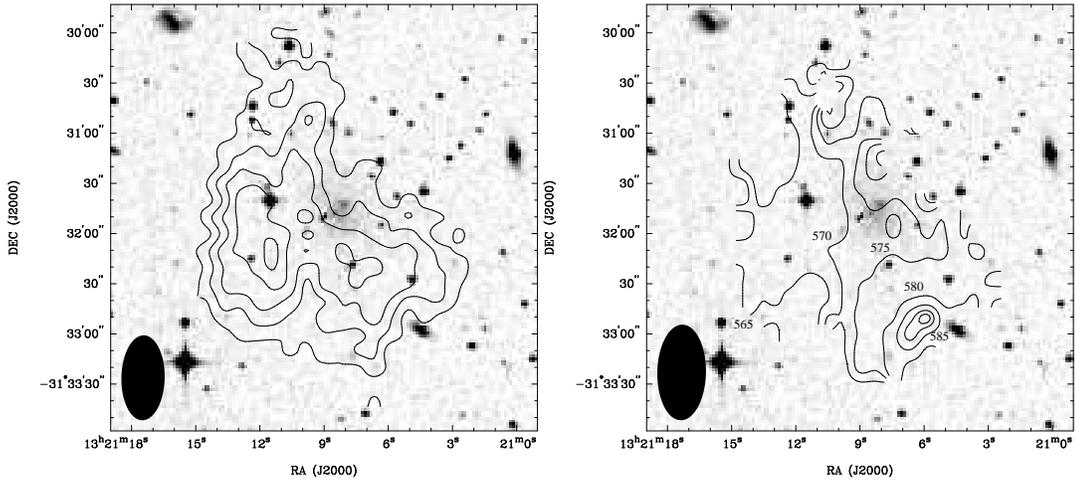

\begin{minipage}{126mm}
\includegraphics[width=63mm,angle=-90]{Figure18a.eps}
\includegraphics[width=63mm,angle=-90]{Figure18b.eps}
%\vspace{63mm}
    \caption{HIPASS J1321-­31: {\em Left:} \hi contour density maps overlaid on the DSS image of the galaxy. The
contour levels are: 0.5, 1.0, 1.5, 2.0, 2.25, 2.35 $\times
10^{20}$ \cmdue. {\em Right:} The \hi velocity field. There is
evidence of a gradient in the velocity field, but it is difficult
to infer from this map clear evidence of organized rotation. The
beam size is shown at the bottom left corner of the image.}
    \label{31HI}
\end{minipage}
\end{figure*}

Despite the significant amount of neutral hydrogen compared to its
stellar mass, there is no evidence of ionised gas in HIDEEP
J1337-­3320 to constrain the current abundance of the ISM, as was possible
with HIPASS J1337-­39. Therefore the age and the metallicity
evolution of this galaxy must be estimated using only the CMD.

We find a dominant population of red giants which, from the shape
of the RGB, comparison with the stellar tracks, and the
distribution of the stars, appears to be of at least intermediate
age.  J1337--3320 {\em seems not} to be a {\em young} galaxy, but
rather has experienced star formation activity over the last 10
Gyr.  This scenario is consistent with the {\scshape starfish}
modelling. The observed large gas fraction appears to be the
result of a very low SFR, which never exceeds $2 \times 10^{-3}$
\msun yr$^{-1}$. After the main event at $10^{+2.5}_{-2.0}$ Gyr
when most of the red giant stars in the galaxy are formed
(according to our model CMD), the subsequent evolution seems to be
characterised by smooth episodes of SF activity and periods of
quiescence.

From the colour of the red giant branch we have obtained a measure
of the metallicity of the intermediate-age population, and we find
 [Fe/H] $= -1.9 \pm 0.6$ which indicates a very low metal content. The synthetic CMD built with
low metallicity isochrones are compatible with that value, but the
best-fit SFHis is given by the combination of three sets of
isochrones (Z = 1/50, Z = 1/20, Z = 1/5 of the solar abundance),
which would give indication of chemical enrichment throughout the
evolution of this dwarf.

An estimate of the metal content can also be inferred from the
metallicity luminosity relation (Skillman et al. 1989, Richer \&
McCall 1995, hereafter RM95). Following RM95 we derive that $12 +$
log(O/H) $= 7.2 \pm 0.2$, where the error is given by the
uncertainty  on the coefficients in the relation of RM95. This
corresponds to a metallicity  in the range between 1/70 and 1/30
of the solar value (with [Fe/O] = 0). The metal content of the
recent stellar population derived from the model SFH ranges
between  Z = 0.001 (1/20 solar) and Z = 0.004 (1/5 solar) (Fig.
\ref{33SFH}), thus with these values the galaxy would stand out in
the luminosity-metallicity relation.

To conclude, the general properties of J1337-­3320, such as the
low luminosity, the spherical symmetry of the optical appearance,
and the absence of current SF activity despite the large gas
reservoir, make this dwarf similar to the so called ``transition
objects'' of the LG (e.g. DDO 210, LGS 3 and Phoenix), which show
intermediate properties between dIrrs and dSphs. These systems are
preferentially found nearer than dIrrs (but not as close as dSphs)
to the most massive members of a group and are generally
interpreted as dIrr galaxies going through a period of temporary
interruption of their star formation activity.

\section{HIPASS J1321--31}

HIPASS J1321-­31 is the most puzzling object among the three. With
a mass of $M_{HI} \simeq 4 \times 10^7$ \msun and an apparent
magnitude $m_B = 17.1 \pm 0.2$ mag (Banks et al. 1999), it has the
highest gas-mass-to-stellar light ratio (\ml $\simeq$ 5), yet
H$\alpha$ snapshots taken with the WIYN telescope do not reveal
any \hii regions. It is also the galaxy with the lowest SB among
the three dwarfs: Banks et al. (1999) found  $\mu^B_0 = 24.2$ mag
\arcsecdue.  It appears to be located in the region surrounding M
83, but its CMD is markedly different from the other dwarfs in our
sample, showing that it has undergone a different SFH (Pritzl et
al. 2003, hereafter Pr03).

\subsection{The neutral gas content}

The \hi distribution of HIPASS J1321-­31 (Fig. \ref{31HI}) extends
out to a radius of 1\arcmin30\arcsec$\,$ (2.3 kpc at the distance
of 5.2 Mpc from the TRGB derived in Pr03) and is clearly offset
from the optical centre. There are two peaks in the \hi density
but they fall near the edge of the chip where the galaxy is
located (WF3), showing poor correlation with the main optical
counterpart.  The column density is low overall and peaks at only
$2.5 \times 10^{20}$ cm$^{-2}$.

A small gradient in the velocity map can be seen in Fig.
\ref{31HI}, but generally the velocity field seems rather
disturbed. It is difficult to infer evidence of rotation from the
contours, however it appears a variation in the kinematics between
the south-western and north-eastern regions of the galaxy.
%%%It is possible to identify a clump of gas in the
%%%south-western area of the galaxy.

The \hi profile  is centred at $579 \pm 1$ \kms , with a 20\%
velocity width $\Delta$$v_{20} = 59$ \kms and a velocity
dispersion $\sigma_v \simeq 8$ \kms. We derive a total \hi mass of
$M_{HI} = 3.7 \times 10^7$ \msun (at a radial distance d = 5.2
$\pm$ 0.3 Mpc).

The dynamical mass, without any inclination angle correction, is
found to be $M_{dyn} = 7 \times 10^8$ \msun, giving
$M_{HI}/M_{dyn} \simeq 0.05$. We have looked through the ATCA cube
to inspect the local environment of the dwarf. There is no sign of
\hi clouds or other low mass gas-rich objects in the 30\arcmin
$\times$ 30\arcmin field of view up to our column density limit of
$N_{HI} = 5 \times 10^{19}$ \cmdue.

\begin{figure}
\includegraphics[width=84mm]{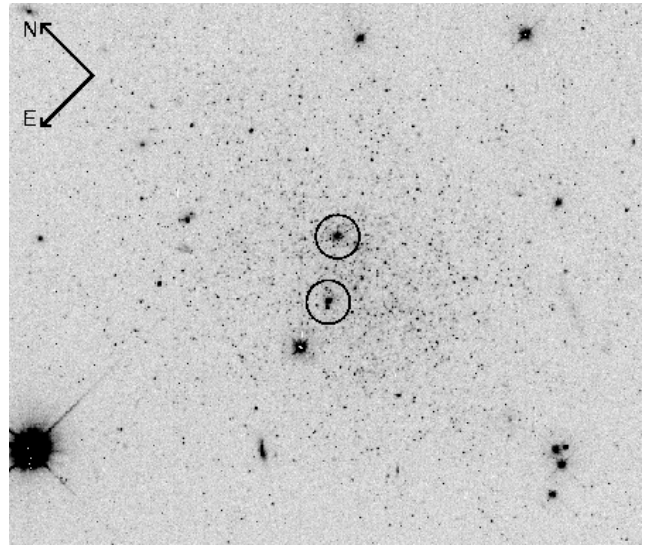}
%\vspace{84mm}
    \caption{WFPC2 image in the F555W filter of HIPASS J1321-­31. The
two circles show the candidate "fading" \hii regions in the
galaxy. The size of the field is approximately 80\arcsec.}
    \label{31redplume}
\end{figure}

\begin{figure}
%\begin{figure*}
%\begin{minipage}{126mm}
\includegraphics[width=87mm]{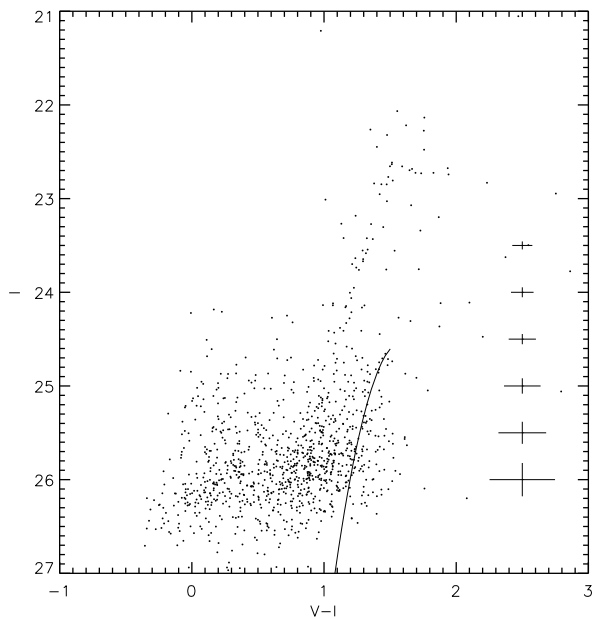}
%\includegraphics[width=63mm]{Figure20b.eps}
%\vspace{63mm}
    \caption{HIPASS J1321-31: ($V - I$, $I$) colour magnitude diagram,
once again compared with the RGB of the GCC M 15, showing a clear
red plume of luminous stars extending up to $I$ = 22.6 mag. }
%%%{\em Right:} CMD of the two fields
%%%around the galaxy. As it appears from the diagrams there is not a
%%%remarkable large population of stars at magnitudes and colours
%%%corresponding to the red plume.}
    \label{31CMD}
%\end{minipage}
%\end{figure*}
\end{figure}

\subsection{The optical properties}

Optical observations of HIPASS J1321--31 (Banks et al. 1999) show
a faint ($m_B = 17.1$), dim ($\mu_0^B = 24.2$ mag \arcsecdue)
galaxy, whose profile is fitted by an exponential law.

A much more detailed view of the dwarf is given by our WFPC2 image
(Fig. \ref{31redplume}), which shows a very diffuse stellar
distribution. There are hardly any lumps or substructures, but one
can see two small, bright regions about 2\arcsec$\,$ across and
7\arcsec$\,$ apart. It is difficult to tell from the image whether
they are \hii regions or star clusters, but they are not
foreground stars. Given the low column density around the optical
counterpart (1 - 2 $\times 10^{20}$ \cmdue), the presence of \hii
regions in this galaxy would be very unusual. If they are  regions
of SF they were too faint to be detected in the H$\alpha$
snapshots from the WIYN telescope.  It is possible that these are
two areas where the SF activity was stronger in the past and which
are now ``fading''.

\begin{figure*}
\begin{minipage}{126mm}
\includegraphics[width=63mm]{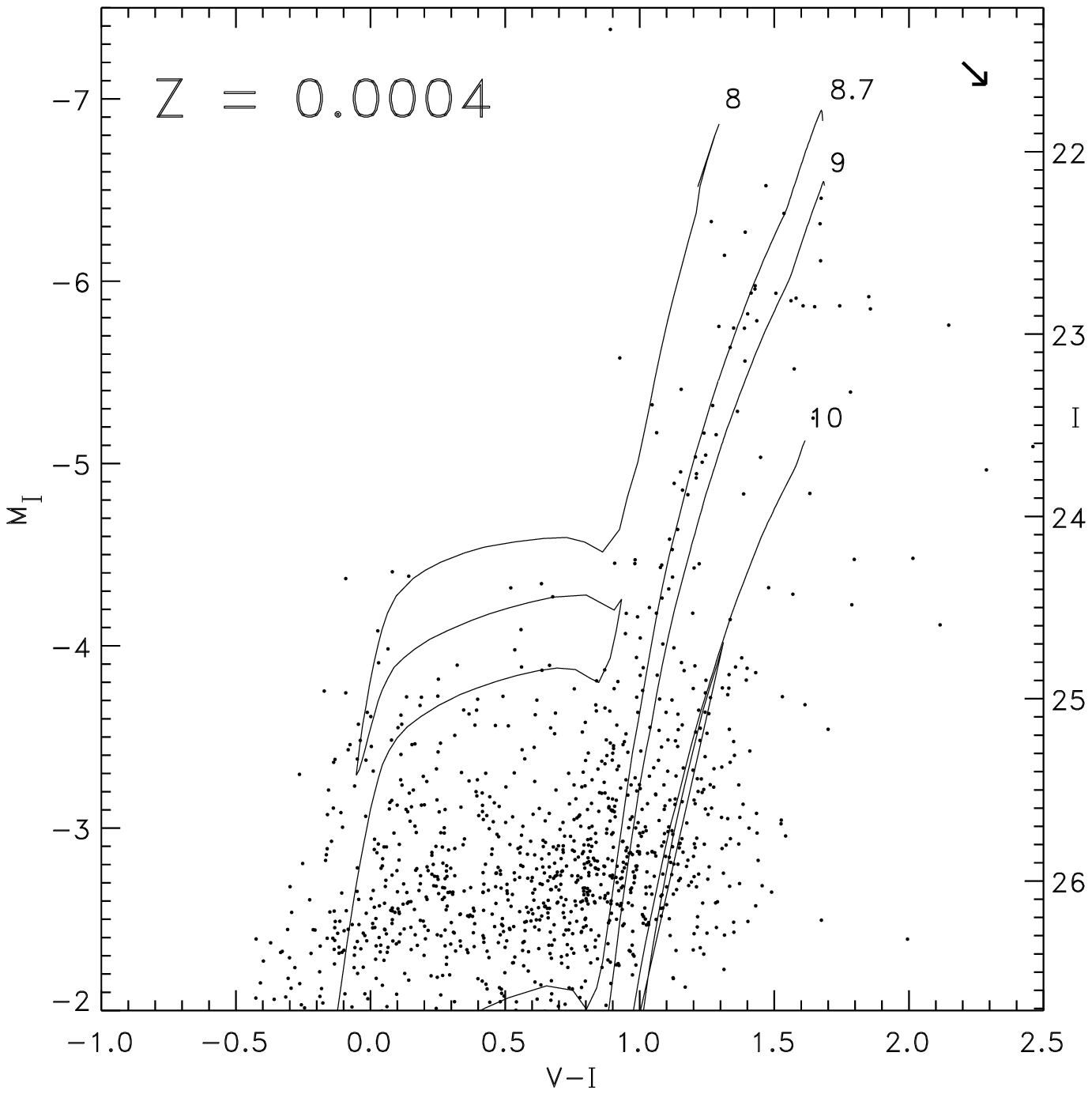}
\includegraphics[width=63mm]{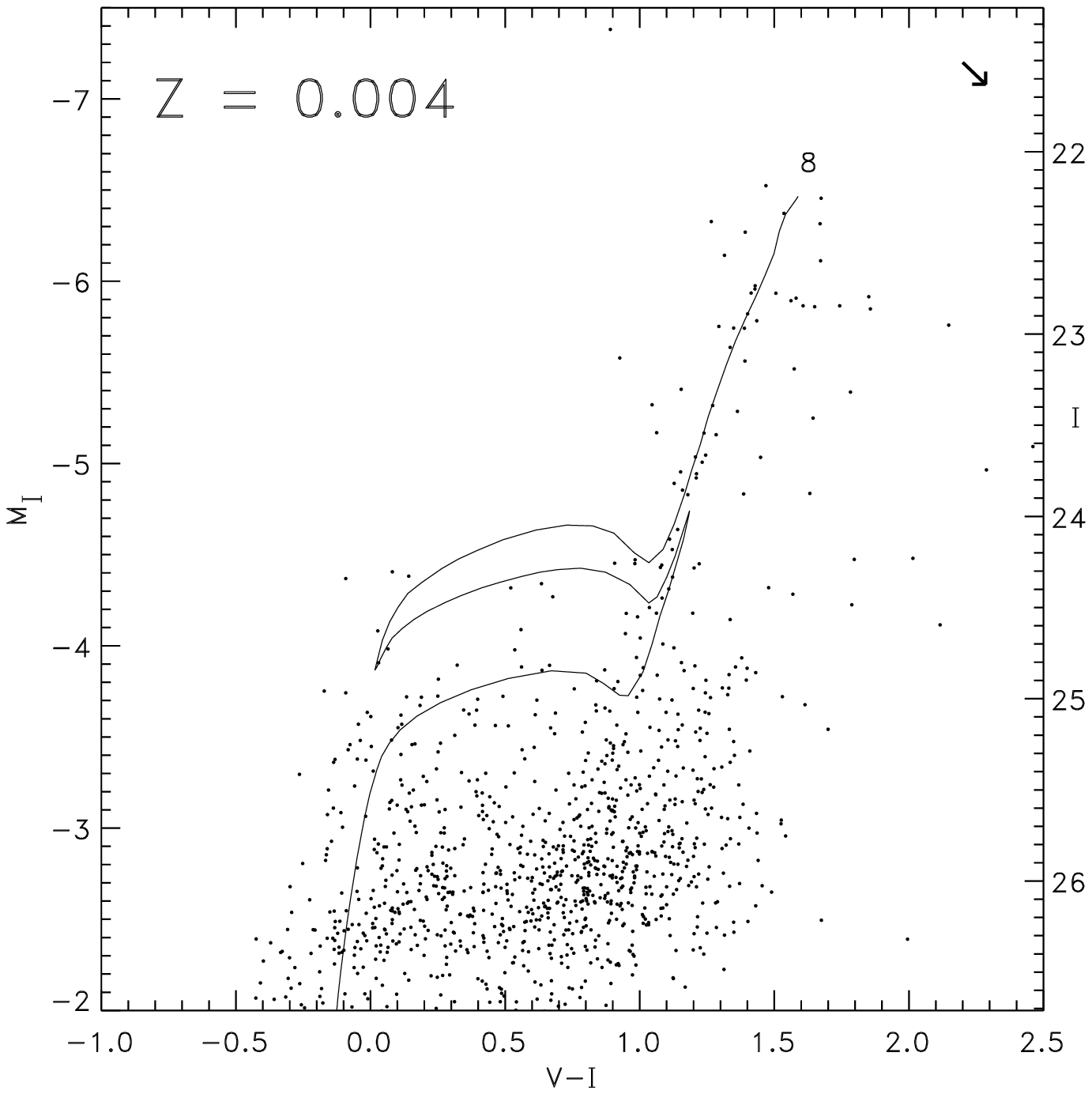}
%\vspace{63mm}
    \caption{HIPASS J1321-­31: dereddened ($V - I$, $I$) colour magnitude
diagrams overlaid on Padua stellar evolutionary tracks at
metallicities of Z = 0.0004 and 0.004. For the highest metallicity
we show both the isochrone that fits the location of the red plume
(to show how the age of the feature may change with the metal
abundance) and the oldest isochrone that would set an upper limit
to the age of the RGB at that metallicity. The arrows at the
top-right corners  indicate the reddening vector.}
    \label{31iso}
\end{minipage}
\end{figure*}

\begin{figure*}
\begin{minipage}{126mm}
\includegraphics[width=63mm]{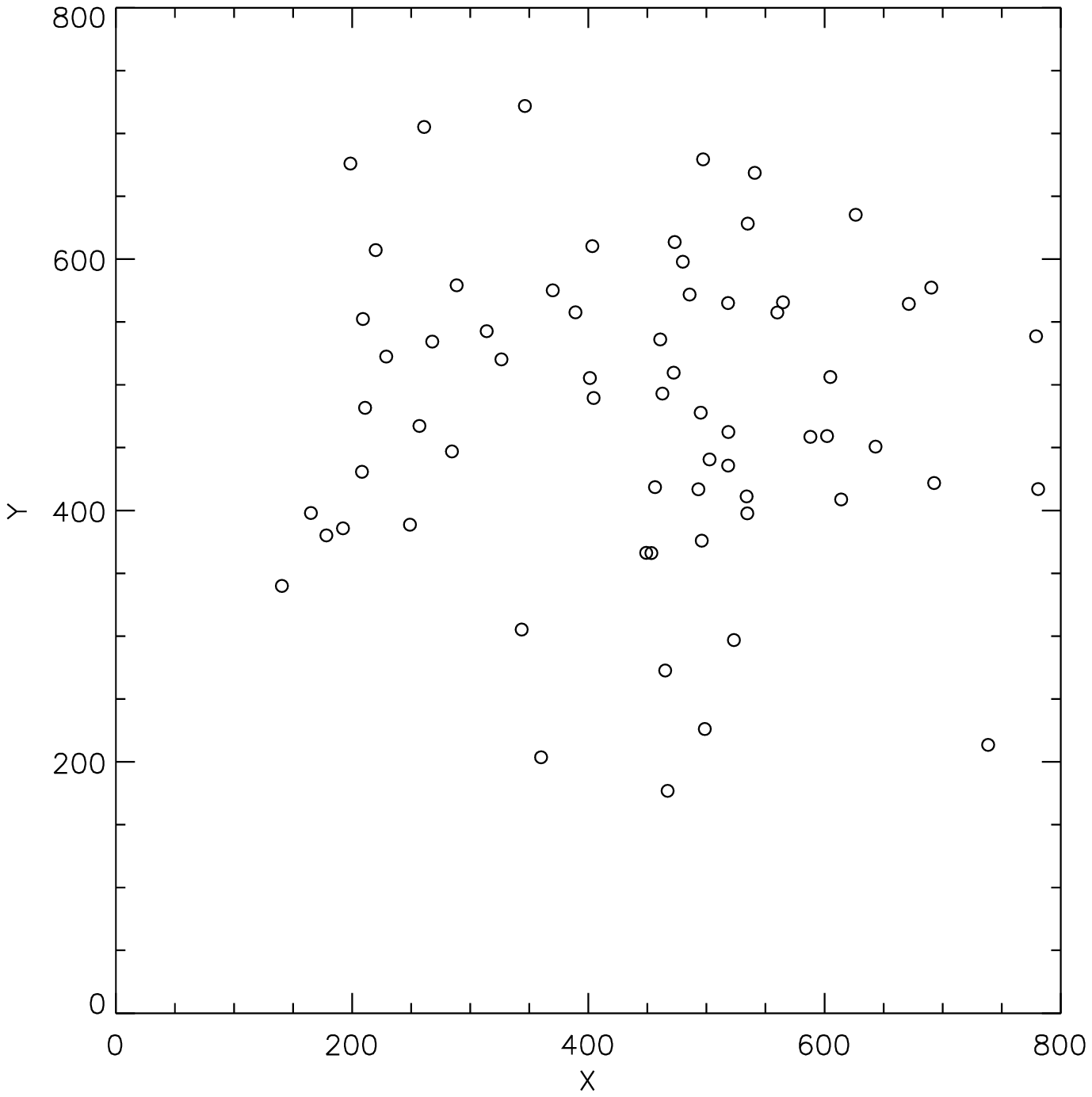}
\includegraphics[width=63mm]{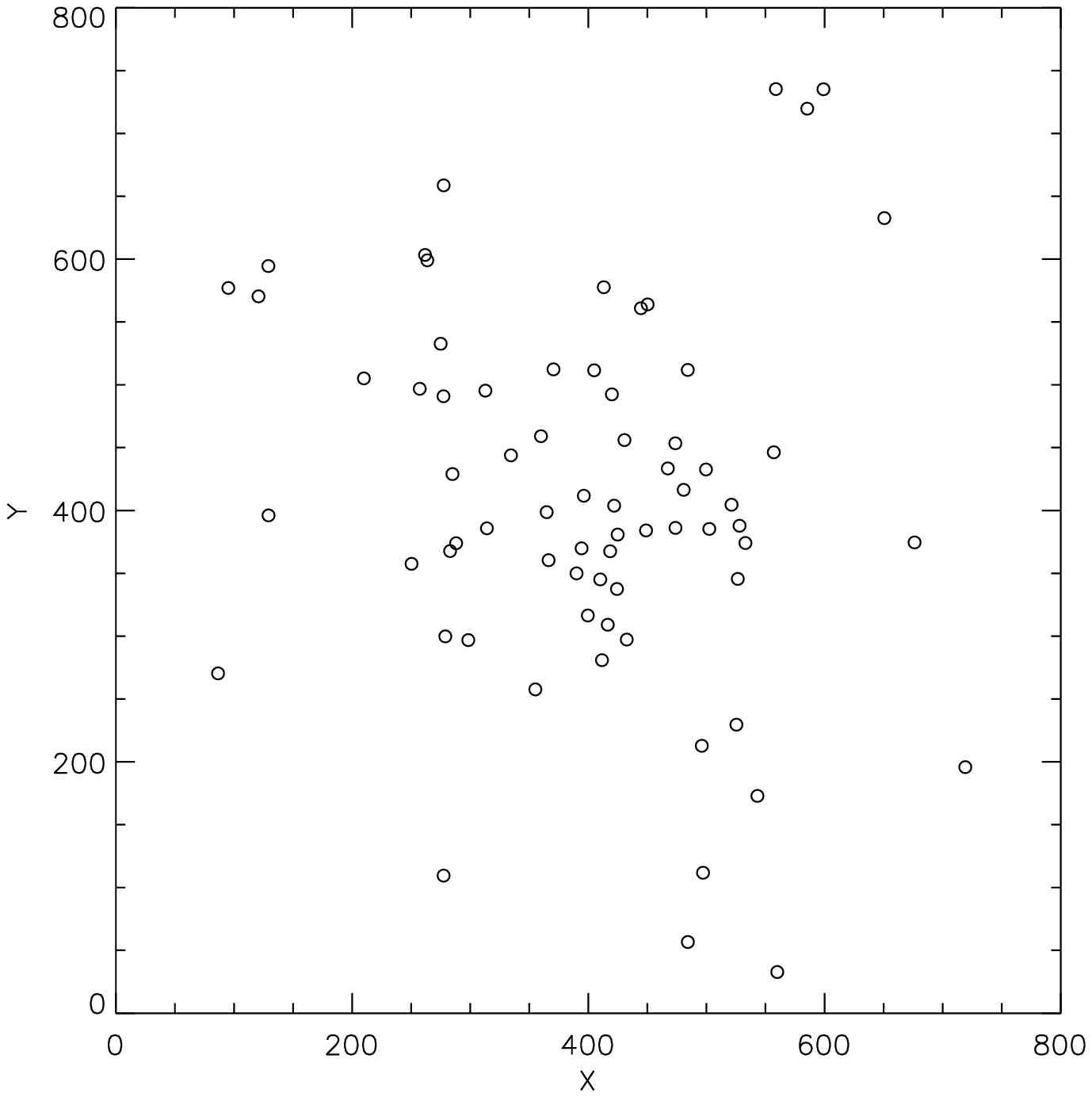}
%\vspace{63mm}
    \caption{The spatial distribution of the stars in the red plume ({\em left}) and of the brightest blue ones with
$I < 25.5$ ({\em right}). There is a difference in the
distribution of the two populations which suggests that they have
been formed at different epochs. Such a difference in location
rules out the hypothesis that the red plume has a higher
metallicity, Z = 0.004, and that these stars are the redder
counterpart of the ones in the blue hump.}
    \label{31bluhigh}
\end{minipage}
\end{figure*}

\subsection{The morphology of the CMD}

The most striking feature of the CMD of HIPASS J1321--31 is the
very unusual red plume which extends at bright magnitudes up to $I
\sim 22.6$ with colours 1 $< (V - I) < 2$ (Fig. \ref{31CMD})
without a corresponding population of bright blue stars. Blue
stars have been detected but they reach at most $I$ = 24 mag .
Among the galaxies of the LG with resolved stellar photometry
there is not another object with a similar CMD. Sextans A for
example (Dohm-Palmer et al. 1997) does show a bright red plume,
but it also has a clear blue plume made of massive stars burning
helium in their core, which are evolving into the bluest extent of
the so-called 'blue-loop' phase.

The bright red stars of J1321-31 are scattered across the optical
extent of the galaxy (Fig. 1 in Pr03), and that seems to indicate
that they belong to the dwarf.

Three possible scenarios have been discussed to interpret the
nature of these unusual red plume stars in Pr03. At first glance
the most obvious option would be that the red plume is the upper
extension of the RGB. If so, the tip would be at $I \simeq 22.6$
mag, a much brighter value than the one expected at the distance
of the Centaurus A group ($I \sim 24$ mag) and this would place
the galaxy in a much closer position to us, at about 2 Mpc. We
have ruled out this possibility because:

\begin{itemize}

\item the heliocentric velocity of the \hi distribution is
compatible with the galaxy membership to the Centaurus group;

\item if that were the RGB its luminosity function between $I =
22.6$ mag and $I \sim 24.5$ mag would be relatively flat, while
the number of stars in the RGB is always found to increase from
the tip down to lower magnitudes, even when it appears narrow and
sparsely populated as in Leo A (Tolstoy et al. 1998).
\end{itemize}

We have also excluded that  the red plume is made of AGB stars
(see Pr03 for details).

We have suggested instead that a more likely scenario for the red
plume is that it is populated by core-helium burning stars in the
Red Super Giant phase.  We can use the comparison with theoretical
isochrones and the simulation of CMDs to further constrain the
interpretation given in Pr03.

The Padua isochrones can be adjusted to the observed magnitudes of
J1321-­31 to set age constraints on the different features. We
have chosen Z = 0.0004 as the lowest metallicity set after
comparing the RGB to that of M 15 (Fig. \ref{31CMD}). The lifetime
of these stars at such metallicity is around 500 Myr (see Fig.
\ref{31iso},) and their mass is in the range 2 - 3 \msun . The
synthetic tracks indicate that the red core helium burning phase
would be much brighter while the corresponding blue loop stars
would be too faint to be detected in our observations. The
presence of the faint blue stars at $-0.1 < V - I < 0.4$ implies
that the SF process must have continued after this event, probably
at a decreasing rate, and dropped off at an age that can be set
around 100 Myr ago (see left panel of Fig. \ref{31iso}).

Adopting a slightly higher metallicity, Z = 0.001 (not shown in
the figure), results only in a variation in the age of the red
plume, but the overall star formation scenario would be the same.
The RSGB stars and those in the blue ``hump'' would have different
ages and correspond to different SF events. The age of the RSGB
stars would be around 300 Myr in this case.

With an even higher metallicity (Z = 0.004, right panel of Fig.
\ref{31iso}) a 100 Myr old isochrone can fit both the brightest
stars of the ``blue hump'' (with $I < 25.5$) and the red plume. In
this scenario, blue and red stars formed at the same time, at a
more recent epoch, and the enhancement in the SF dates back to
only 100 Myr. However this interpretation seems very unlikely
because blue and red plume stars show different spatial
distributions: the first are more centrally concentrated (see
right panel of Fig. \ref{31bluhigh}) while the latter appear
randomly distributed throughout the optical body of the galaxy
(Fig. \ref{31bluhigh}, left panel). If both type of stars were
coeval it would be difficult to explain their different location
within the galaxy. Thus red plume stars have to be older than the
blue ones to explain their scatter and we can rule out the higher
metallicity (Z = 0.004) scenario.

The constraint on the metallicity of the red plume and the
comparison with stellar tracks is the only tool we have at this
stage of the analysis to estimate the age of the RGB, that we have
defined as the feature located at $1.0 < V - I < 1.7$, extending
up to $I \sim 24.5$. If we assume that the metallicity of the
plume is Z = 0.0004 - 0.001, we derive the usual scenario of a
very low metallicity galaxy with an RGB which is fitted by
isochrones as old as 10 Gyr (Fig. \ref{31iso}, left panel). The
TRGB has been found at $I_0 = 24.43\pm 0.11$, assuming $E(B-V)=
0.06 \pm 0.01$, with a distance modulus $(m -M)_0 = 28.59 \pm
0.13$,  that corresponds to a distance of 5.2 $\pm$ 0.3 Mpc
(Pr03).

\begin{figure}
\includegraphics[width=84mm]{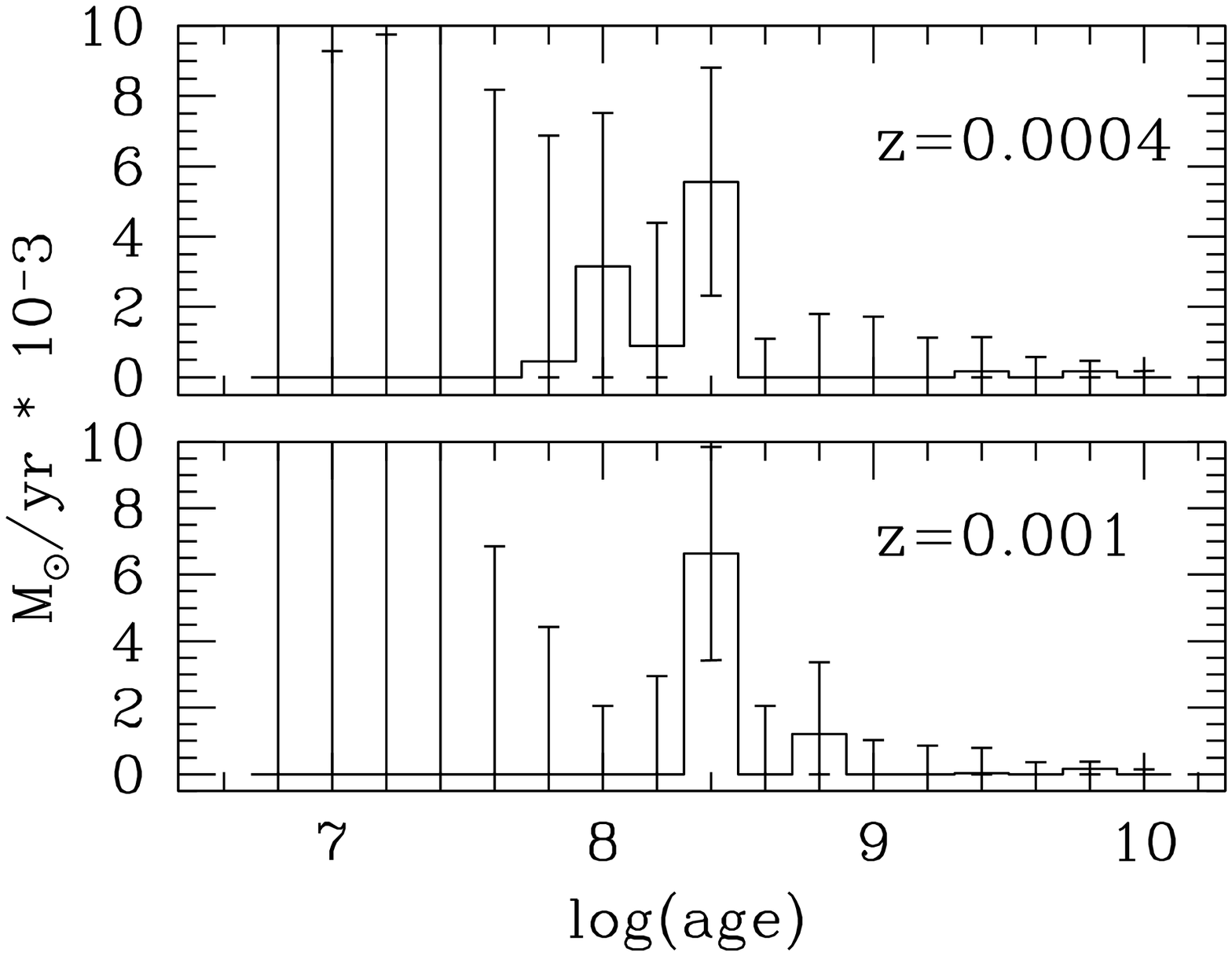}
%\vspace{84mm}
    \caption{The best-fit SFH of HIPASS J1321--31}
    \label{31SFH}
\end{figure}

\subsubsection{A bursty SFH?}

\begin{table}
\caption{The list of the different combination of metallicities
inspected with {\scshape starfish} and the corresponding reduced
$\chi^2$ parameter. For HIPASS J1321-31 we find that the best-fit
SFH is given by a model with isochrones at Z = 0.0004 and 0.001.}
\begin{center}
\begin{tabular}{|ccc|}
  \hline \hline
  % after \\: \hline or \cline{col1-col2} \cline{col3-col4} ...
  Z &  & $\chi^2$ \\
  \hline \hline
  0.0004 &  & 1.56\\
  0.001 &  &  1.41\\
  0.004 &   & 1.47\\
  0.0004+0.001 & & 1.30  \\
  0.001+0.004 &  & 1.40 \\
  0.0004+0.001+0.004 & & 1.34 \\
  \hline \hline
\end{tabular}
\end{center}
\end{table}

\begin{figure*}
\begin{minipage}{126mm}
\includegraphics[width=63mm]{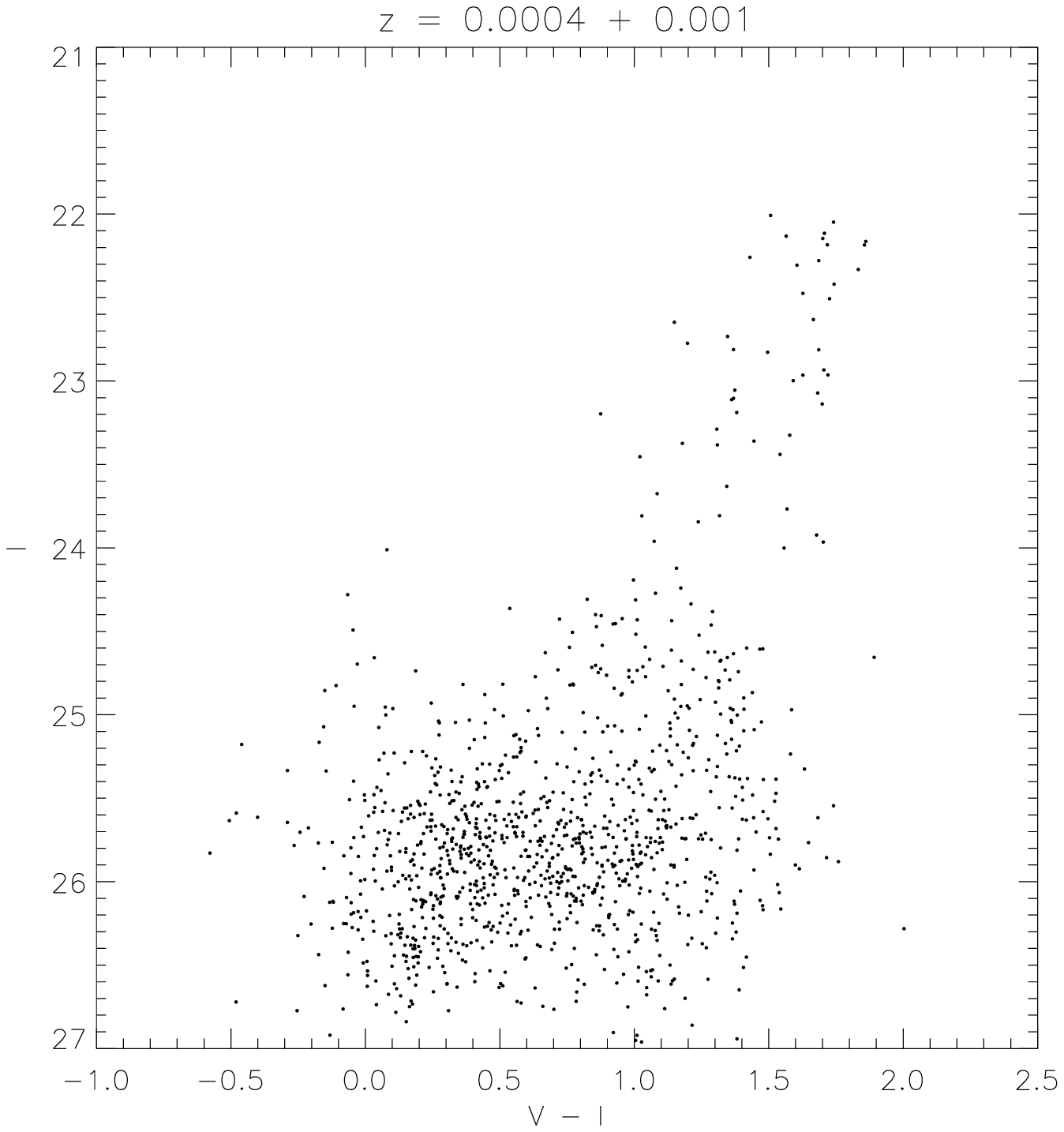}
\includegraphics[width=63mm]{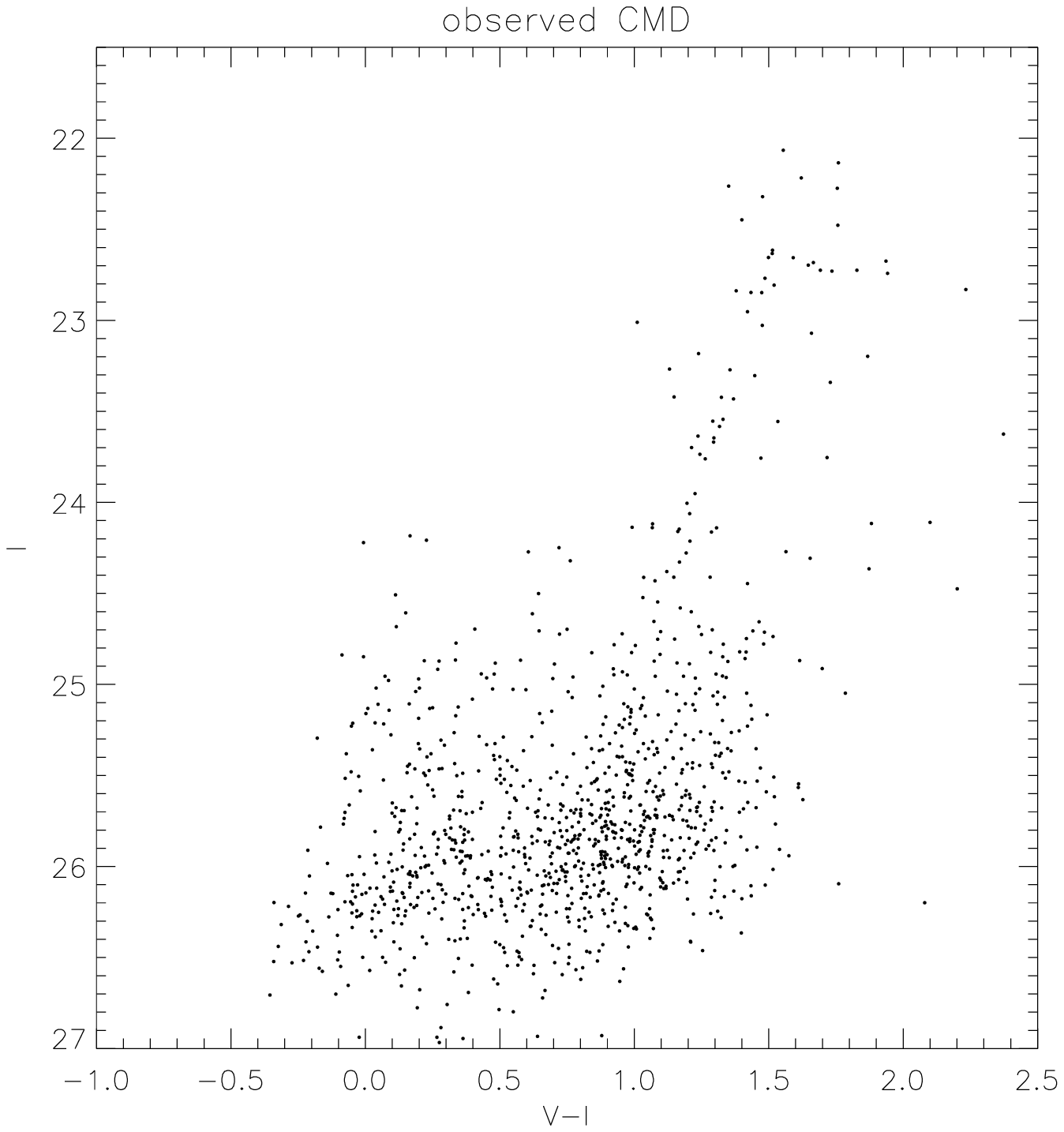}
%\vspace{63mm}
\caption{The best-fit model CMD ({\em left})  of HIPASS J1321--31
obtained using the set of isochrones with Z = 0.0004, 0.0001  is
compared to the observed one ({\em right}).} \label{31CMDsynt}
\end{minipage}
\end{figure*}

We have modelled the SFH and CMD of J1321-­31 with {\scshape
starfish}, trying different combinations of three sets of
isochrones (Z = 0.0004, 0.001, 0.004). The resulting fits obtained
for different models are listed in Table 5. Using tracks with Z =
0.0004 and Z = 0.001 yields the best-fit SFH (Fig. \ref{31SFH})
and the model CMD (Fig. \ref{31CMDsynt}) for comparison with the
observed CMD. We find that a model CMD with metallicity as high as
Z = 0.004 cannot reproduce the observed data.

The best-fit SFH suggests that J1321-­31 formed its red giant
branch at $\log(Age) = 9.8 \pm 0.1$, i.e. the first detectable SF
event occurred at $6^{+1.6}_{-1.3}$ Gyr ago. The RGB appears to be
relatively ``young''. This does not exclude the existence
of a population of (fainter) stars older than 10 Gyr that we
cannot detect, such as HB stars. Also, it appears likely that
the algorithm has underestimated the older SFR of the galaxy: the model
RGB is less populated than the observed one.
There may be some problems in modelling the RGB as the galaxy is more
distant than the other dwarfs, thus the stars are closer to the
photometric limit and have larger photometric errors.

With regard to more recent SFH, the red plume produced by the code
is significantly wider than that observed.  In the simulation, the
feature begins to appear at $\log(Age) = 8.8$, but the majority of
the stars form at $\log(Age) = 8.4$. We would expect the narrow
red plume to be the result of a single burst of SF at a single
metallicity. However the small number of stars in it made it
difficult for the code to model such a sharp burst.

At $\log(Age) = 8.4$ the SFR is about $6 \times 10^{-3} \msun$
yr$^{-1}$, six - ten times higher than the past SF episodes
(taking into account the uncertainties on the SFRs). If we use
this value to estimate the mass in stars born in the age bin at
$\log(Age) = 8.4$, we obtain $M_{\star} \simeq 7 \times 10^5$
\msun , which is rather high compared to a luminosity of $7.4
\times 10^6$ \lsun. It is possible that the recent SFR is slightly
overestimated because the number of predicted blue stars is higher
than those observed, nevertheless {\scshape starfish} finds that a
large fraction of the stellar population was formed in the recent
past, and that the galaxy probably went through a ``bursty" phase.

On the other hand the enhancement of the SFR that occurred in
J1321-­31 is not as intense as those found in BCDs. In VII Zw 403
for example, the SFR at its peak was $\sim$ 30 times higher than
normal, and the starburst phase lasted for a few hundred million
years (Schulte-Ladbeck 1999).

In summary, the simulations of the SFH of J1321-­31 have given us
the following insights into the SFH of J1321-­31:

\begin{itemize}

\item Low metallicity isochrones provide a better
fit, supporting the hypothesis that the galaxy has an
intermediate-age stellar population. As we go back to ages older
than 1 Gyr we find that the RGB stars are only about
$6^{+1.6}_{-1.3}$ Gyr old.

\item The SFR seems to have been significantly higher in the last
500 Myr than it was between 1 and 10 Gyr ago. The red plume
originated around 500 Myr ago although the theoretical feature is
not as narrow as the one observed.

\end{itemize}

\subsection{Summary and discussion}

HIPASS J1321-­31 appears to be a very uncommon galaxy both from
its gaseous distribution and optical morphology. It has a large
amount of \hi (with the highest \ml ratio among the three
objects), with a very irregular distribution and low column
density ( $\lesssim 2.5 \times 10^{20}$ \cmdue ) which is clearly
offset from the optical counterpart.

The stellar distribution is very diffuse and is characterised by
bright red stars which are scattered throughout the optical area
of the galaxy and form a peculiar red plume when plotted on the
CMD. The lack of a corresponding population of bright blue stars
indicates that the galaxy has experienced a peculiar star
formation history. Three scenarios for the origin of this
anomalous red plume were discussed in a previous paper (Pr03) -
RGB, AGB, or RSGB stars - and we concluded that it is most
probably composed of core helium burning post main sequence stars
with ages less than 1 Gyr which are likely to be related to an
epoch when the galaxy experienced an increase in the SFR.

To verify this interpretation we have built models of the SFH of
the galaxy with different metallicities and found that the best
fit was given by assuming a very low abundance for the stellar
population (between 1/50 and 1/20 solar) implying, as in the other
dwarfs, that the red giant branch is made of intermediate-age
stars ($< 10$ Gyr).

The SFH between 1 and 10 Gyr is characterised by very few events
and an overall low SFR (below $2 \times 10^{-3}$ \msun $yr^{-1}$).
We find that the RGB stars are only $6^{+1.6}_{-1.3}$ Gyr old, a
few Gyr younger than what we have estimated for the other two
dwarfs. Such a low rate is consistent with the scenario of a
gas-rich object evolving ``slowly'', compared to the large amount
of gas present (van Zee et al. 1997a).

At more recent ages ($<$ 1 Gyr) the model SFH confirms that the
red plume is related to an epoch of increased SFR as we suggested
in Pr03.  The predicted SFR implies that a large fraction of the
stellar population of J1321-­31 was produced during this event.
The offset of the gas distribution and the interruption of the
star formation activity may provide further evidence that the
galaxy went through a phase of enhanced SF activity, which heated
and pushed away the gas from the optical centre of the galaxy.
After this main episode the SFR started to decrease and the SF
activity stopped around 100 Myr ago.

As we have done with J1337-3320, an estimate of the metal
abundance from the luminosity-metallicity relation may be useful
as a comparison to the values  used in our simulations. Following
Richer \& McCall (1995) we obtain 12 + log(O/H) $= 7.4 \pm 0.2$,
or 1/45  $<$ Z $<$  1/15 of the solar value(again with [Fe/O] =
0). In this case the range of metal abundances obtained from the
CMD analysis  would be compatible with the luminosity-metallicity
relation.

HIPASS J1321-­31 appears  different from the other two dwarfs. It
is not a young galaxy, formed less than 1 Gyr ago, since we have
detected the RGB and have established an age of 6 Gyr for its
oldest red giant stars. However it appears to have a predominantly
young population with an age less than 500 Myr.

\section{Discussion}

%%%%There are several issues that we have tried to answer with this

The high \ml ratios and low optical luminosity of these dwarfs
make them unique objects in the nearby universe. Despite  large
fractional \hi mass they have not used the majority of their gas,
and  do not show obvious signs of massive SF. We have tried to
understand why.

The first issue was to determine if we had found a class of young
galaxies, the most natural explanation for their high
gas-mass-to-stellar-light ratios. In all the three dwarfs
examined, a clear RGB has been detected. There is a  difference
between the CMDs of these dwarfs and that of I Zw 18, the nearest
young galaxy detected so far, with a stellar population that seems
to be less than 500 Gyr old, in which a well identified RGB is
missing (Izotov \& Thuan 2004). The finding of  red giant stars in
these dwarfs sets a lower limit on their age of at least 1 - 2
Gyr, the minimum time for such stars to appear after a first SF
episode. In principle, even much older populations (10 - 12 Gyr)
can be found in  the RGB, and the presence of a well defined tip,
as we have observed in  the three dwarfs, is generally considered
as a reliable proof for the existence of stars older than 1 - 2
Gyr
(Tolstoy et al. 1998). %%This is the first evidence of  a several
%%Gyr population.

Secondly, we have looked for AGB stars whose detection would
provide unambiguous evidence of  an intermediate-age population (1
- 10 Gyr) (Caldwell et al. 1998). Two out of three galaxies
(J1337--39 and J1337--3320) do show candidate AGB stars. The lack
of AGB stars in J1321--31 is puzzling since we would expect to
find such population given that the RGB has been well detected.

Modelling has offered another point of view on the issue. Using
the {\scshape starfish} code we have found that SF activity
between 8 and $\sim$ 12 Gyr ago  would explain the appearance of
the RGBs in both J1337--39 and J1337--3320. These dwarfs may host
older populations of stars which are beyond our detection limits.

The situation is less clear for J1321--31. Its first episode of SF
%(in the $\sim$ 10 Gyr time range we have considered)
seems to have occurred at $6.3^{+1.6}_{-1.3}$ Gyr ago. The model
SFH shows  a very weak SFR until about 1 Gyr ago and only in the
last 300 Myr we find a significant increase in the SF activity,
which then drops again in the last 100 Myr.

Therefore  these galaxies are not recently formed objects, and
their evolution has proceeded  for several Gyrs, most likely with
modest episodes of SF activity alternating with periods of very
low production of stars or even quiescence.

From their model  SFHs we can estimate the average SFR, and we
obtain over a 10 Gyr time range the following values: $2 \times
10^{-3}$ \msun yr$^{-1}$ (J1337--39), 0.6 $\times 10^{-3}$ \msun
yr$^{-1}$ (J1321--31) and 0.2 $\times 10^{-3}$ \msun yr$^{-1}$
(J1337--3320).

The \hi distributions appear similar to what it is seen in dwarf
irregular galaxies (e.g. van Zee et al. 1997b; Young et al. 2003).
The neutral gas extend well beyond the optical image, and the peak
column density exceeds $ 10^{21}$ \cmdue only in one case, where
there is also a rough correspondence between the \hi density peak
and regions of current star formation.

The gas depletion time scale $\tau_g \equiv M_{HI}$/SFR gives an
indication of the time needed to consume the present amount of gas
at a given rate. We obtain timescales much longer than the Hubble
time ($\sim$ 20, 60, 25 Gyr respectively), therefore these
galaxies should be able to retain much neutral hydrogen for
several more gigayears.

Another estimate of the formation time scale of these dwarfs is
given by  their specific star formation rates, i.e. per unit mass
of stars, the inverse of the time it would take to form the
present stellar mass of a galaxy at a given SFR. The
``galaxy-building timescales" for these dwarfs are comparable to
the Hubble time (6, 17, 11 Gyr for J1337--39, J1337--3320 and
J1321--31 respectively)\footnote{Starburst systems for example
have typical time scales of 1-2 Gyr, which in certain
cases may go down to 0.1-1 Gyr %as it is found in some very intense
%%and compact ultraviolet sources, the so-called ultraviolet
%luminous galaxies (UVLGs)
(Heckman et al. 2005).}.

We next tried to understand what inhibits the process of SF in
these objects, leaving most of the gas un-processed. Is it a
consequence of their internal properties -- i.e. a too diffuse ISM
with low \hi column densities, and perhaps the absence of a cold
phase of the ISM, with low metallicity and low dust -- or is it
due to their local environment?

\subsection{The role of the gas density and the metallicity of the ISM}

Recent work on the gas distribution in dwarf galaxies indicates
that their gas densities can be well below the threshold for star
formation (Toomre 1964) across the entire stellar disc, suggesting
that it is difficult to initiate a global burst of star formation
(Hunter \& Gallagher 1986; Hunter \& Plummer 1996; van Zee et al.
1997b,c; Hunter, Elmegreen, \& Baker 1998). The local nature of
star formation in dIrr galaxies has been emphasized by Hunter \&
Gallagher (1986). van Zee (1997b,c; 2001a,b) concludes, after the
analysis of several gas-rich dIrr galaxies, that the star
formation process in such objects randomly diffuses across the
stellar disc, occurring only in those localities where the gas
density is sufficiently high to allow molecular cloud formation
and subsequent star formation. According to van Zee star formation
is a localised phenomenon
%%%%. The
%%%%onset of the star formation activity across the galaxy is expected
%%%%to be a quiescent phenomenon since
and it is unlikely that the entire disc will have sufficient gas
density to permit a global starburst.
%%%%van Zee has also compared the gaseous distribution of
%%%%gas-rich LSB dIrrs and BCDs and found a fundamental difference
%%%%between the two types of galaxy.
Quiescent LSB gas-rich dwarfs show a shallow \hi distribution,
roughly constant throughout the disc and below $10^{21}$
cm$^{-2}$, in contrast to BCDs where the gas is centrally
concentrated, and the density peaks are above 10$^{21}$ cm$^{-2}$
(van Zee 2001a).

In our small sample, current SF occurs only in J1337--39, the
galaxy where the \hi density is highest and above $10^{21}$
cm$^{-2}$. This is also the galaxy with the highest average SFR.
The other two systems, where the peak column density is about one
order of magnitude lower ($N_{HI} \lesssim$ 2 $\times 10^{20}$
cm$^{-2}$) and the gaseous distribution is almost constant
throughout the optical disc, show no signature of current SF.

%%%%%We can get an estimate of the SF threshold for these objects by
%%%%%calculating the critical density, $\Sigma_c$, using the Toomre
%%%%%criterion
%%%%%(1964). %%%, although the validity of such a relation for low mass and
%%%%%%%%irregular galaxies is still under debate.
%%%%%Assuming a thin, isothermal disc,  the critical gas density is
%%%%%given by
%%%%%
%%%%%
%%%%%\begin{equation}
%%%%%%%%\Sigma_c = \frac{\alpha c}{3.36 G}\left[1.4 \frac{V}{R} \left( 1 +
%%%%%%%%\frac{R}{V} \frac{dV}{dR} \right)^{\frac{1}{2}} \right] \Sigma_c =
%%%%%\Sigma_c = \alpha \frac{c_s k}{\pi G}
%%%%%\end{equation}
%%%%%
%%%%%where $c_s$ is the velocity dispersion in the gas, %%%which ranges
%%%%%%%%between 6 - 9 \kms,
%%%%%$\alpha$ is a dimensionless geometrical parameter with a value of
%%%%%0.7 for spiral galaxies (Kennicutt 1989), $k$ the epicyclic
%%%%%frequency defined as $k^2 = 2 (\frac{V}{R})^2 \left[ 1 +
%%%%%\frac{R}{V} \frac{dV}{dR} \right]$, $R$ is the radius of the
%%%%%galaxy, $V$ the rotational velocity at the radius $R$. In our
%%%%%simplified approach we will assume a solid body
%%%%%%%%%rotation curve (V $\propto$ R).
%%%%% rotation curve (V $\propto$ R), and we
%%%%%find that the critical column density for J1337--39 is N$_{HI}^c$
%%%%%$\sim 1 \times 10^{21}$ cm$^{-2}$, and its central column density
%%%%%exceeds such a critical value only close to \hii regions. %Similar
%%%%%%calculations for the other galaxies give similar values, while
%%%%%%they have almost constant \hi column density below N$_{HI}^c$.

%%%The lack of high star formation rates and starbursts is probably
%%%due to

The low \hi surface densities and the low metal content imply that
the large amount of gas available for SF may be in a warm phase (T
$> 100$ K), preventing the transition from atomic to molecular
hydrogen that cools a gas cloud and leads to its fragmentation. In
fact, according to Elmegreen and Parravano (1994) the transition
from a warm ($T \sim 10^4$) to a cold phase ($T \sim 100$ K) of
the ISM is essential to make the gas gravitationally unstable and
trigger SF (see also Elmegreen 2002). %%%Skillman et al. (1987) found that SF in dwarf
%%%irregulars occurs at a column density of $5 \times 10^{20}$
%%%\cmdue. Elemgreen (2002) suggested that this column density
%%%corresponds to the minimum pressure necessary
%%%%%%
%%%%%%It is possible that quiescent gas-rich galaxies do not experience
%%%%%%intense SF events because their gaseous discs are globally stable
%%%%%%against the gravitational collapse because their ISM is in a warm
%%%%%%phase ($T \sim 10^4$).
Schaye (2004) argues that the phase transition in conditions of
local hydrostatic equilibrium, assuming  a single phase warm ISM,
depends only on the density (and column density) of the gas if the
temperature, gas fraction, metallicity and UV radiation field are
fixed. SF will occur in those regions where the gas surface
density exceeds the local threshold value. Such a critical density
correspond to the critical pressure necessary for the existence of
a cool diffuse phase. %%%even if SF is  on average
%%%suppressed for these objects, because their surface density is in
%%%general lower than the critical value,
In his model, SF can occur  locally only where the pressure is
high enough to trigger the phase transition from a warm to a cold
medium (Schaye 2004). According to Schaye, at this threshold
pressure the surface density is $\Sigma_c \sim 3 - 10 \msun$
pc$^{-2}$ (or $N_{H,crit} \sim 3 - 10 \times 10^{20}$ cm$^{-2}$).

The temperature of the ISM is related to the cooling rate of the
gas, which depends on its metal content. The radiative
cooling-time of the gas heated by supernovae  defines the
timescale in which it falls back into the galaxy, where it can
eventually be reprocessed in a new SF event. According to
Hirashita (2000) the cooling time is given by

\begin{equation}
t_{cool} \sim 7 \times 10^8 \left( \frac{\zeta}{0.1} \right)^{-1}
yr \label{Hirashita}
\end{equation}

where $\zeta$ is the metallicity of the galaxy normalized by the
solar abundance. According to Hirashita irregular SF activity is
common in small size dIrr galaxies and is the consequence of the
balance between the heating process of the ISM due to SN, stellar
winds and the emission of UV radiation (i.e. stellar feedback) and
the cooling rate.
%%%%For a characteristic radius of 1 kpc the heating
%%%%from the stellar feedback is very efficient with a characteristic
%%%%propagation time scale of $10^8$ yr where a propagation speed of
%%%%the ''feedback wave'' of $\sim$ 36 \kms has been assumed following
%%%%Dopita et al. (1985).
Thus the combination of  small size (which makes the heating from
stellar feedback  more effective) and low abundance (determining
an inefficient cooling rate) conspire to produce an irregular and
``intermittent'' SFR in small-size dIrr galaxies  (Hirashita
2000).

For our sample of galaxies the metallicity is around 1/20 - 1/50
solar, therefore, from Eq. \ref{Hirashita},  the cooling times
should be between 1.5 and 3.5 Gyr. Therefore their low gas
densities and low metal abundances may explain the low SFRs and
their ``gasping'' SFHs.

\subsection{The efficiency of SN-driven mass and metal ejection}

Establishing that the SF in these dwarfs has occurred for several
Gyr raises two more issues. First we should check whether the
energy injected into the ISM by SNe and stellar winds, given the
derived SFRs, can affect the HI content of these galaxies.
Secondly, how can SF activity extended so long in time be
consistent with the low metal abundances we have estimated?

Recent models, which include a large dark matter component, have
investigated the range of parameters that contribute to the
removal of the ISM in a galaxy after an enhanced SF activity, and
they show that it is much more difficult to remove the ISM with a
single starburst than previously thought (Mac Low \& Ferrara 1999,
hereafter McF99).

The evolution of a galaxy after an intense SF activity is
predicted by McF99 for different gaseous and stellar masses, and
for various mechanical luminosities of the starburst $L$, i.e. the
fraction of the total energy released by a supernova into the ISM
as mechanical energy, given a certain rate of SNe per year. This
parameter is expressed as $L_{38}$ (units of $10^{38}$ ergs
s$^{-1}$).
%Determining the efficiency of mass loss implies the
%introduction of some assumptions on the nature of the dark matter
%%%halo.
%%%The dark-to-visible mass ratio used in the simulations corresponds
%%%to $\phi = M_h / M_{bar} = 34.7 (M_{bar}/10^7$ \msun)$^{-0.29}$
%%%(Persic et al. 1996), where M$_h$ stands for the mass of the dark
%%%matter halo.
According to McF99 mass ejection (with the
disruption of the gaseous content) is very efficient only for
galaxies with M$_{gas+stars} = M_{bar} \lesssim 10^6$ \msun.
Galaxies with M$_{bar} \gtrsim 10^7$ \msun are less affected by
starburst events.

\begin{figure}
\begin{center}
\includegraphics[width=84mm]{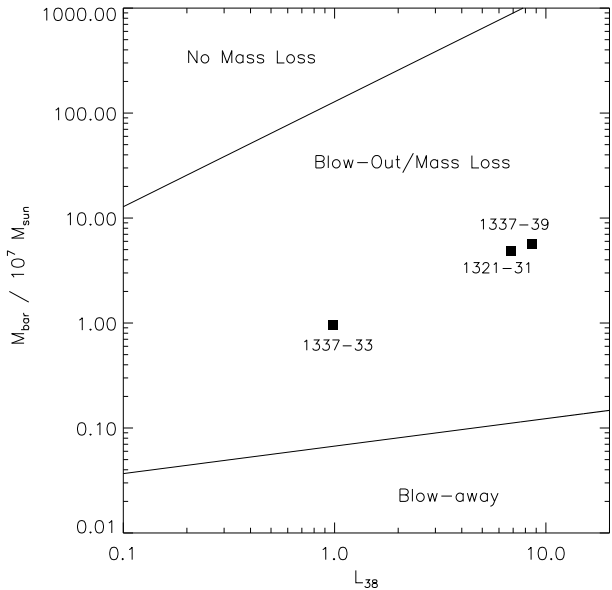}
\caption{Values of the baryonic mass of a dwarf galaxy and the
starburst luminosity showing the regions where blow-away, blow-out
or no gas loss may occur (see McF99 for details). The filled
squares represent the location on the plane of the HIPASS dwarfs,
having considered the most intense SF events in their recent
history. As one can see from the picture, all the dwarfs galaxies
are in the area of the diagram where only blow-out may occur, i.e.
only the hot enriched gas is ejected. The episodes of SF are not
able to eject the whole ISM and the galaxies according to the
McF99 model are able to retain the bulk of their gas content.}
\label{Blow-out}
\end{center}
\end{figure}

To have an idea of the effects on the dwarfs of the most intense
SF episodes obtained with {\scshape starfish}, we have calculated
the L$_{38}$ parameter for each galaxy as in  MCF99. The baryonic
mass of our dwarf galaxies ranges between 10$^7$ \msun $\lesssim
M_{bar} < 10^8$ \msun. Using the dark-to-visible mass ratio ($\phi
= M_h / M_{bar}$, where M$_h$ stands for the mass of the dark
matter halo) given by Persic et al. (1996) we obtain 35 $ \gtrsim
\phi \gtrsim$ 20\footnote{Note that this is a factor 2 higher than
the dynamical-baryonic-mass ratios found from the velocity widths
of the 21-cm emission lines (20 $\gtrsim M_{dyn}/M_{bar} \gtrsim$
10)}.  We have assumed that the energy released by a SN event is
about $E_0 = 10^{51}$ ergs, a tenth of which is deposited into the
ISM while the rest is radiated away (Thornton et al. 1998) We have
estimated the amount of stellar mass produced in the SF episodes
with the highest SFR for each dwarf in their recent past where the
time resolution of the SFH is higher (for example for J1321--31 we
have considered the event at log(Age) = 8.4, for J1337--39 the
most recent event at 10 Myr, and for J1337--3320 the one at
log(Age) = 7.8). We have then calculated the number of SNe
produced in each SF event ($N_{SN}$), using a Salpeter mass
function, and derived L$_{38}$ as $N_{SN} E_0 / 10 \Delta t$ where
$\Delta t$ is given by the width of the corresponding age bin. We
have then plotted the results in the M$_{bar}$ - L$_{38}$ plane as
defined in McF 99 (Fig. \ref{Blow-out}). %%%%%%In the plane three main
%%%%%%areas are defined: the {\em Blow-Away} region, where the ISM is
%%%%%%completely ejected from the galaxy, the {\em Blow-Out/Mass Loss}
%%%%%%region which defines the masses and mechanical luminosities at
%%%%%%which only part of the ISM is ejected from the galaxy during a
%%%%%%starburst, and finally the upper-left corner region defines where
%%%%%%the mass losses are negligible ({\em No Mass Loss}).

As one can see from the figure, all the three dwarfs are included
in the area of the plane labelled as {\em Blow-Out/Mass Loss}
where they undergo blow-out of the gas, but where the ISM will not
be completely ejected.

According to the simulations of McF99, the effect of the stellar
winds and SNe on the metal abundances is much more dramatic. About
60\% of the metal-enriched material is expelled from a galaxy with
M$_{bar} = 10^9$ \msun, a fraction that increases up to 100\% for
M$_{bar} = 10^8$ \msun, which is the order of magnitude of the
total baryonic mass of our dwarfs. Metals easily escape from low
mass dwarf galaxies because the chemically enriched material
produced by the stars is entrained within the hot bubble of gas
where it reaches the edge of the galaxy and the sound speed is
higher than the escape velocity (Legrand 2000).
%%%%%When this bubble
%%%%%reaches the outer envelope of
%%%%cold gas, the hot gas inside blows right away leaving the cold
%%%%material behind.
%%%%%%%In the more massive objects (M$_{bar} > 10^9$ \msun ) a
%%%%%%%significant fraction of the metal-enriched gas is retained by the
%%%%%%%gravitational potential of the dark matter halo and will
%%%%%%%eventually falls back onto the galaxy after it cools down (McF
%%%%%%%99).

On the other hand, if a low mass dwarf galaxy experiences low
luminosity SF events, with  a lower rate of SNe, the outflow of
the metal-enriched gas into the intergalactic medium may be
reduced (McF99; Ferrara \& Tolstoy 2000) leading to a gradual
increase in the metal abundance of its ISM. This is what may have
happened in J1337--3320, which is the galaxy with a higher
metallicity among the three dwarfs, and a lower average SFR, and
it seems to show two stellar populations with
different chemical abundance. %%%From analysis of the CMD,  the RGB
%%%and the very red AGB stars cannot be fitted by the same set of
%%%evolutionary tracks, suggesting that
Its ISM  may have been chemically enriched by the older generation
of RGB stars and the metals may have been reprocessed into  the
later SF events.

\subsection{The effects of the environment}

M 83 is the closest massive spiral to the HIPASS dwarfs, and we
have calculated their location and dynamical status in relation to
this bright galaxy to check if there is a possible connection
between the environment of the dwarfs and their observed
properties. The distance from M83 is assumed to be

\begin{equation}
R^2 = D^2 + D^2_{M83} - 2D \cdot D_{M83} cos \theta
\end{equation}

where $\theta$ is the angular separation from M 83, D the distance
of each dwarf derived from its TRGB and D$_{M83}= 4.5$ Mpc.

%%%%%The radial velocity relative to M83 can be calculated by
%%%%%
%%%%%\begin{equation}
%%%%%(V - V_{M83}) = V \cdot cos \lambda - V_{M83} \cdot cos(\theta +
%%%%%\lambda)
%%%%%\end{equation}
%%%%%
%%%%%where $\tan \lambda = D_{M83} \cdot \sin \theta/[D - D_{M83} \cdot
%%%%%cos(\theta)]$ (Karachentsev et al. 2002).
%%%%%%In Tab. 7.1 we display
%%%%%%the results for the three objects.

Excluding  J1337--3320 ($R \sim 300$ kpc), the location of the
other two dwarfs in the group seems  similar to those of  dIrrs in
the LG such as WLM, SagDIG or Gr 8, at about 850  kpc (J1337--39)
and 800 kpc (J1321--31) from the more massive galaxies of
the group. %%%%Moreover the three dwarfs all present positive radial
%%%%velocities relative to M83.

Being in the outskirts of the group
%%%%the lack of close massive
%%%%%%galaxies or gas clouds may have prevented
the triggering of frequent star formation events through
collisions or tides in the discs of  J1337--39 and J1321--31 may
not have occurred. This could be advanced as an additional
explanation for their 'retarded' evolution.
%% so that they have
%%been able to preserve most of their primordial gas content.
Because of its relative closeness to M 83, the case of
J1337--3320, which indeed has the lowest \ml of the three, is
slightly different.

Gas-rich dIrr galaxies are found in the LG, but the main
difference from the HIPASS dwarfs is that they are not as
gas-rich. Their \ml ratios in fact are in general around or below
1. We may wonder  if one of the reasons why we do not find this
type of dwarf galaxy in the LG is because of an intrinsic
difference between the two environments.

\begin{figure}
\begin{center}
\includegraphics[width=84mm]{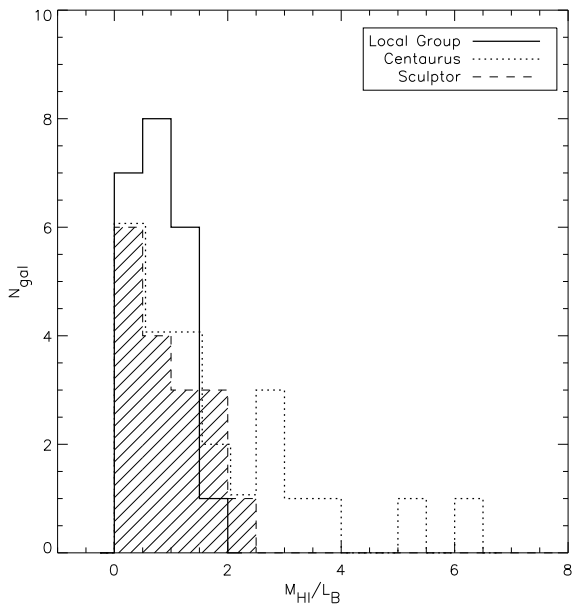}
\caption{The comparison between the \ml ratios of 24 galaxies in
the Centaurus A group ({\em dotted line}), 22 galaxies in the LG
({\em solid line}), and 17 galaxies in Sculptor ({\em dashed
line}). As one can see the majority of the galaxies in the LG have
ratios less than 1.5, while the sample of the Centaurus A and
Sculptor galaxies extend to higher \ml ratios.} \label{LG_vs_CenA}
\end{center}
\end{figure}

In Fig. \ref{LG_vs_CenA}  we compare the gas fractions of the
smaller members of the LG, Sculptor  and Centaurus A as a whole
group\footnote{For those members for which distances are not
known, it is not possible to distinguish whether they belong to
the Cen A or M 83 subgroup. Therefore for this comparison we
consider Centaurus A as a single group},
 without
including their primary galaxies.  The \hi sources in Centaurus A
have been taken from C\^ot\'e (1997), Banks (1999), and
Karachentsev et al. (2002), and the updated fluxes have been
extracted by the HIPASS brightest galaxy catalogue (Koribalski et
al. 2004). The reddening corrected optical magnitudes have been
taken from the Nasa Extragalactic Database (NED). The \ml ratios
for the dwarfs of the LG  come from Mateo (1998). For Sculptor the
objects have been taken from  SCM03, Karachentsev et al. (2003)
and Bouchard et al. (2005). Some of the objects with the higher
\ml ratios in SCM03, such as Sc 18 and Sc 24, have revealed to be
false detections (Karachentsev et al. 2003) and have not been
included in the list. For other objects with high \ml ratios such
values have been recalculated using the more recent data from the
HIPASS brightest galaxy catalogue and HOPCAT, the optical
counterparts for HI radio- selected galaxies (Doyle et al. 2005),
giving smaller values than those in SCM03.

As one can see in the Figure, 95\% of the satellites of the LG
have \ml ratios below 1.5 in solar units, against $\sim$ 60\% of
the objects in Cen A. The remaining 40\% have much higher \ml
ratios up to $\sim$ 6 (see also Knezek 1999). The higher average
number of gas-rich satellites does not seem to be the consequence
of the different crossing time. Comparing the subgroups formed by
the Milky Way and M83, Karachentsev et al. (2002) find that the
velocity dispersion of the the M83 satellites correspond to  a
crossing time of  2.3 Gyr, while in the Milky Way it is 2.1 Gyr
(Karachentsev 2005). Graham (1976) and van Gorkom et al. (1990)
tried to explain this issue suggesting that the Centaurus A
complex has recently accreted a population of gas-rich dwarfs, but
the velocity dispersion of its members, typical of groups of
galaxies of this size, does not seem to justify this
interpretation (Knezek 1999). Moreover, the group is well known to
be gas-rich at all luminosities (Knezek 1999; Banks et al. 1999),
since even its early-type systems such as NGC 5102 and NGC 5128
are very gas-rich for their types. Therefore there does not seem
to be an obvious explanation to justify the difference in the gas
content between the Centaurus A complex and the LG. Sculptor also
contains gas-rich objects (about the 25\% of the smaller galaxies
detected at 21-cm have \ml $> 1.5$). The  group is known to be a
loose aggregation of galaxies -- extending for about 6 Mpc along
the line of sight -- and it does not seem to be a a virialized
system given its long crossing time -- 6.9 Gyr for the bigger
subgroup centered around NGC 253 (Karachentsev et al. 2003).
Therefore, interaction between galaxies are expected to be low and
external gas stripping largely ineffective in Sculptor, yet the
gass-mass-to-light  ratios are not as high as those found in
CentaurusA/M83.

\subsection{Comparison with other dwarf and irregular galaxies}

\subsubsection{Similar objects within the LG}

The stellar populations of most  dwarfs in the LG have been
studied in detail, therefore the  comparison with these systems
may give further hints on the understanding of our gas-rich
dwarfs.

Sag DIG is probably the galaxy that most resembles J1337--39 and
it is gas-rich with \ml $= 1.6$. It is considered the most remote
object belonging to the LG, since it is located at a distance of
1.29 $\pm$ 0.09 Mpc from the LG barycenter (van den Bergh 2000b).
From its RGB Sag DIG appears the most metal poor dIrr of the LG
(\fe = --2.1 $\pm$ 0.2) (Momany et al. 2002).  The youngest stars
are located near the major peaks of emission in the \hi
distribution (which appears as a complete asymmetric ring with a
central depression), whereas the red giants and intermediate-age
AGB stars define an extended halo or disc with scale length
comparable to the size of the hydrogen cloud. The CMD of this
fairly gas-rich dwarf obtained by Momany and collaborators (2002)
 has many features in common with that of J1337--39: the well
populated blue plume, the blue RGB typical of a low metallicity
population, and the evidence for AGB stars above the TRGB.
%%%%%   shows a very low metallicity RGB population (\fe = --2.1
%%%%%$\pm$ 0.2, the lowest in the LG), an increase in the recent SF
%%%%%activity between 30 and 100 Myr, and signs of ongoing formation of
%%%%%stars, $\dot{M}_{SAG} = 0.95 \times 10^{-3}$ \msun \yr \kpcdue
%%%%%(Hunter \& Elmegreen 2004). Its current SFR is comparable to
%%%%%1337-39 ($\dot{M}_{39} =  1.5 \cdot 10^{-3}$ \msun \yr \kpcdue).
%%%%%%%%% or disc
%%%%with scale length comparable to the size of the hydrogen cloud.
%%The photometric data available for Sag DIG to date are not deep
%%enough to infer the presence of an older stellar population
%%(Saviane et al. 2002).
Deeper observations carried out with the Advanced Camera for
Surveys (ACS) on the HST (Momany et al. 2005) show also the
presence of a red HB, proving that SF started around 10 Gyr ago in
this galaxy.

Therefore dIrr galaxies in the LG do contain an ancient ($\gtrsim
10$ Gyr) stellar population. HB stars have been detected also in
WLM ($M_B= -13.9$; \ml = 1.2; Mateo 1998) and it is the only dwarf
that contains a globular cluster (Hodge et al. 1999). The time of
initial SF has been set at about 12 Gyr ago (Dolphin 2000b), again
showing that isolated and fairly gas-rich dIrr galaxies of the LG
have experienced SF even at ancient epochs.

The general properties of the remaining two galaxies, J1337--3320
and J1321--31, such as the low luminosity, the spherical symmetry
of the optical appearance, the absence of current SF activity
despite the large gas reservoir, make these dwarfs more similar to
the so called ''transition objects'' of the LG such as LGS 3,
Antlia, Pegasus (Miller et al. 2001; Gallagher et al. 1998; Lee et
al. 1999). All these galaxies have low luminosities ($M_B$ =
$-9.9$; $-10.2$; $-12.3$) and \ml less than 1 in solar units. They
are among the lowest mass objects ($M_{\star} \gtrsim 10^6$ \msun)
which have been able to retain gas and to experience SF until a
few hundred million years ago and they show how the process of gas
consumption and quiescence takes place in dwarf galaxies. Moreover
they are within 300 - 400 kpc from either the Milky Way or M 31.

According to Gallagher et al. (1998), SF in Pegasus (\ml = 0.2) is
temporarily ``down'', but it is not excluded that it could
re-establish a normal level of SFR in future. The prominent RGB
and the extended AGB indicate that the SF at intermediate-ages was
more intense than during its more recent history.
%%%%shows a current low SFR
%%%%$\approx 3 \times 10^{-4}$ \msun yr$^{-1}$. If its evolution
%%%%continues to proceed at this rate, it will take about 13 Gyr to
%%%%exhaust all its gas.  Pegasus would have been brighter about 2 Gyr ago
%%%(Gallagher et al. 1998).
This galaxy may be taken as an example of ``gasping" SFH, since it
appears to have experienced occasional intense episodes of SF in
its past, lasting for 0.5 - 1 Gyr, followed by more quiescent
phases, lasting one to few Gyr (Cole et al. 1999).

LGS 3, in the M 31 subgroup, with its smooth gaseous and stellar
distribution is probably the galaxy that most resembles
J1337--3320. Its SFH appears not to have been characterised by
strong bursty events. After an early event around 13 - 15 Gyr it
has been followed by a uniform and low SFR (around 10$^{-4}$ \msun
yr$^{-1}$) (Miller et al. 2001).
The smooth chemical enrichment of
the stellar population suggests that there have not been
sufficiently violent bursts of SF to expel a large fraction of the
metals, which instead have been reprocessed into stars.

Finding a galaxy in the LG with properties similar to J1321--31 is
difficult, but its relative isolation and recent drop in the SF
activity,  make  DDO 210 a good candidate.
%%%%% shows a SFH similar to J1321-31.
DDO 210 shows a low luminosity ($M_B = -10.95$) and is fairly
gas-rich (\ml = 0.87) (Young et al. 2003).  Its current SFR is
negligible, but it shows an enhancement in the last few hundred
million years compared to the average value for its entire
lifetime (Lee et al. 1999).
%%%The normalized SFR of DDO 210
%%%for 1 $< t <$15 Gyr is compared in Table 7.4 with the SFR of
%%%1321-31 for 1 $< t <$ 10 Gyr.
There is evidence of SF until 30 Myr ago (Lee et al. 1999),
however the recent drop of SF is puzzling because it contains
large amounts of cold neutral gas (at low velocity dispersion of 3
- 5 \kms) with peak density as high as $10^{21}$ \cmdue (roughly
one order of magnitude higher than J1321--31) (Young et al. 2003).

\subsubsection{Comparison with gas-rich galaxies outside the Local Group}

Beyond our Local Group, Sculptor is the closest system of galaxies
hosting  few objects with high \ml ratios (Cot\^e et al. 1997),
whose SF properties have been recently studied by SCM03. Their
SFR, normalised to the galaxy luminosity or gas mass, indicates
that these dwarfs are experiencing modest SF activity compared to
the LG objects, and they show  long gas depletion time scales (in
some cases $\tau_{gas} >$ 100 Gyr) (SCM03). However among the new
gas-rich galaxies discovered by Cot\^e et al. (1997), those with
the most extreme properties (very low SFRs and $\tau_{gas} \sim
1000$) such as Sc 18 and Sc 24 turned out to be false detections
(Karachentsev et al. 2003). The low density of the Sculptor
environment, where interactions between galaxies are expected to
be minimal and external gas stripping largely ineffective, is
probably the most plausible reason to explain the properties of
these galaxies.
 As mentioned in sec 6.3,
the group contains also three 'transition' (dSph/dIrr) galaxies,
which have fainter luminosities compared to the other gas-rich
objects in the group and they show no sign of current SF activity.

\begin{figure}
\begin{center}
\includegraphics[width=84mm]{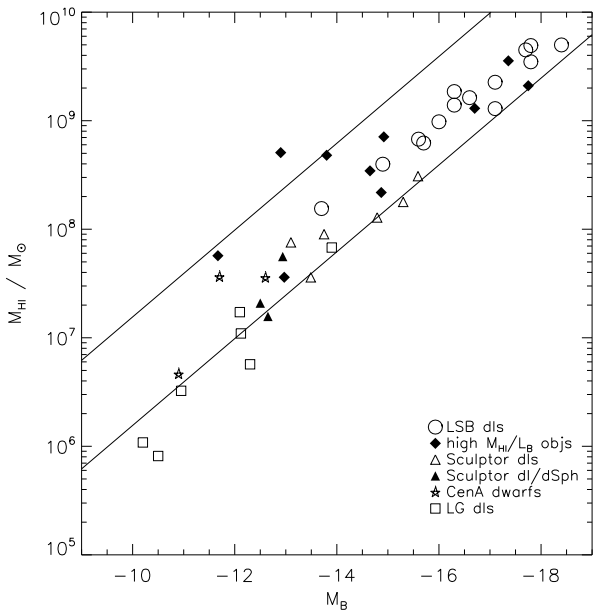}
\caption{ The B absolute magnitude and the \hi mass of the HIPASS
dwarfs ({\em stars}) are compared to the LG dIrrs ({\em open
squares}), the Sculptor dIrrs ({\em open triangles}) and
transition-type galaxies ({\em filled triangles}), the sample of
high \ml objects from Warren et al. (2006) including also  DDO 154
(\ml = 9.4) and UGC292 (\ml = 7) ({\em filled diamonds}), and the
isolated LSB galaxies from van Zee (1997b) ({\em big circles}).
The lines correspond to \ml = 1, 10.} \label{gasrich_compare}
\end{center}
\end{figure}

The Sculptor gas-rich galaxies tend towards larger gas fractions
and lower SFR as the  three dwarfs we have analysed, while the
dIrrs in the LG are not unusually gas-rich and show higher
relative SFRs (SCM03). Nevertheless  the Sculptor dIrrs    cover a
higher range of luminosities when compared to the HIPASS dwarfs
(Fig. \ref{gasrich_compare}). The properties of our sample are on
average more similar to the Sculptor transition type objects, even
though they seem to have  different spatial distribution within
the groups and larger \ml ratios.

A sample of high \hi mass-to-light ratio galaxies has been
recently studied by Warren et al. (2006), obtaining more accurate
measurements of the \ml parameter. Warren and collaborators show
that the estimates of this ratio based on magnitudes in available
catalogues often turn out to be too high. After the new optical
observations only three out of nine galaxies have \ml $> 3$,
including one extremely gas-rich object, ESO 215-G?009. The sample
does not include only dIrrs but also different morphological types
of galaxies much brighter and with a high gas content than the
dwarfs we have studied. However it is interesting to underline
some properties in common with our sample such as their isolation
in space, the  extended \hi discs and the low density of the gas,
factors that as we have discussed in the previous sections prevent
a vigorous SF activity.

In Fig. \ref{gasrich_compare} we compare  the B absolute magnitude
and the \hi mass of the HIPASS dwarfs to the dIrrs of the LG (\S
6.4.2) and the gas-rich objects we discussed in this section, i.e.
the Sculptor galaxies and the sample of  Warren et al. (2006). We
also show the isolated LSB galaxies from van Zee (1997b)
characterised by high \ml ratios and long gas depletion time
scales. However, as one can see from the figure, this sample
covers a range of properties (luminosity, gas masses and also
optical morphology) which is remarkably different from the dwarfs
in Centaurus A.

%%%%%%

\section{Conclusions}

Nearby dwarf galaxies are one of the best environments in which to
study in detail how stellar evolution proceeds. Accurate analysis
of their resolved stellar photometry (when feasible) and of their
gas content (if present) is essential for a better understanding
of the internal processes that influence the SF evolution of a
galaxy, such as the metal abundance and the density of the ISM, or
external effects, such tidal interaction or merging.

 Dwarf galaxies in the LG all show varied and very
complex SFHs. Dwarfs are all unique objects: it is difficult to
find a common pattern in their evolution. It appears that SF has
occurred either continuously over long period of time or, in a few
objects such as Carina or Sculptor, in distinct episodes (Grebel
1997). All LG galaxies for which sufficiently deep photometry is
available do seem to contain at least an intermediate-age
population of stars, and for the majority of them old populations
of stars have been uncovered. Therefore SF started in the LG at an
epoch comparable to the Hubble time.

But what do we know about other groups? Do dwarf galaxies outside
the LG behave in the same way? Have they formed at the same time
as in the LG? In order to get a complete overview of the evolution
of dwarfs we need to extend the same analysis to nearby
environments.

This was one of the aims of this study even though the
observational limits are challenging. Studies of dwarf galaxies
outside the LG have been already carried out, but they have been
mostly restricted to objects with different properties from our
sample, such as luminous BCDs going through an intense period of
SF activity (Schulte­Ladbeck et al. 1999; 2001; Annibali et al.
2003; Cannon et al. 2003; Izotov \& Thuan 2004). Instead the
HIPASS dwarfs with their large gas fractions, their low masses,
SBs and faint luminosities, give an insight into a different
evolutionary environment, where galaxies can evolve in a smooth
and more quiescent way.

While we do find signs of  an intermediate-age population, showing
that  SF occurred several Gyr ago, much deeper data is needed to
verify that the SF activity started in these galaxies at the same
time as in the LG dwarfs, and yet the detection of such stars
would be very challenging.

The idea that very small galaxies should experience a single
primary and possibly cataclysmic episode of SF during their
lifetimes does not appear to fit  the HIPASS dwarfs we have
studied, in analogy with what is found in the LG. We have pointed
out that according to recent simulations (MCF99; Ferrara \&
Tolstoy 2000), their ISMs seem to be stable to SF events even if
they happen to be fairly strong, as we have found in J1321--31.
The HIPASS dwarfs show that low mass galaxies can retain their ISM
even if they have experienced SF activity for several Gyr. We have
also shown that if their evolution were to proceed at the same
average rate that they have experienced so far, it will take at
least another Hubble time to consume all their gas.

Therefore it seems unlikely that these objects will become dSphs.
Their evolutionary rate is too slow and their positions within the
group too isolated
%\footnote{Nevertheless we cannot completely
%exclude that their gas content could be removed in a stripping
%event if
%%%%%passed too close to M83, but their current position within the
%%%%%group do Unless
%their orbits are highly eccentric and at some stage they pass near
%their giant companion M83.}
to reach the SFR necessary to burn all
the hydrogen. With a hydrogen mass of about 10$^7$ \msun they
would need to reach a SFR of 10$^{-2}$ \msun yr$^{-1}$ and to
maintain it for 1 Gyr in order to consume their gaseous fuel. This
seems highly unlikely given their SFHs. Moreover the low gaseous
density, and the inefficiency of the cooling mechanism due to the
low metallicity would prevent the possibility of a long SF period
at a high rate. Shorter episodes of SF activity would require
higher
rates of 10$^{-1}$ \msun yr$^{-1}$ for 100 Myr. %%%%We cannot exclude
%%%%that their \hi could be removed in a stripping event if they
%%%%passed too close to M83, but their current position within the
%%%%group do

This is in agreement with what is found in the LG, where gas
removal in dwarf spheroidals appears to be the combination of
internal processes  (a relatively rapid initial and
intermediate-age SF activity that reduced the gas supplies) and
external factors (the stripping of the depleted gas disc by a
close encounter with a giant galaxy)  (Grebel, Gallagher \&
Harbeck, 2003).

Is it possible for these galaxies to go through a starburst phase
turning into BCDs? One of the key question of dwarf galaxy
evolution is whether there is a link between starbursting systems
and quiescent gas-rich galaxies, whether they are the same objects
viewed at different stages of their evolution. Gas-rich galaxies
in principle have large reservoir of gas to ignite intense SF
activity. However according to van Zee the majority of gas-rich
dIrr do not go through a starburst phase. This is mainly the
consequence of different intrinsic properties between gas-rich
dIrrs and BCDs (van Zee 2001b). BCDs have steeper rotation curves
and lower angular momenta, which lead to centrally concentrated,
high density gas distribution favouring the triggering of a
starburst. From the simulation of the SFH we have found only in
HIPASS J1321--31 a significant increase in its past SFR (between
six an ten times the SFR from 1 to 10 Gyr, taking into account the
uncertainties on the SFR values). It is difficult to say whether
J1321--31 has experienced a BCD phase or not, although the
enhancement in the SF is probably not
strong enough.  %%%%%to argue that J1321--31 went through such a stage.
It seems unlikely that the HIPASS dwarfs went through a starburst
phase in their past, suggesting that factors other than richness
in gas have to play role in the triggering of a BCD phase in dwarf
galaxies.

We are still left with the intriguing issue of what prevents the
conversion of the gas content of a galaxy into stars. We have
suggested some possible mechanisms that may be responsible for the
observed {\em ``retarded''} evolution of the dwarfs found in the
Centaurus A group. Understanding the characteristics of the ISM
and the conditions of the local environment may be crucial in
order to infer why SF is inefficient in these dwarfs. Our analysis
of the resolved stellar populations and the gaseous distributions
goes in this direction and allows us to examine the correlation
between neutral gas and star forming regions, and to estimate the
timescale of the main evolutionary episodes of the history of
these galaxies, as well as their location within the group.
However a complete understanding of what prevents stars from
forming in these environments is still missing.

The next stage of this analysis is  to extend the study of the
stellar populations of gas-rich galaxies to a much larger and more
varied sample. A first obvious target would be the Sculptor group
where similarly high gas-mass-to-stellar-light ratios have been
measured in few objects (SCM03), with the advantage of being
closer than Centaurus A. The analysis of another group would allow
to investigate how different environmental conditions can be
related to the evolution of gas-rich galaxies.

\section*{acknowledgements}
We would like to thank Erwin de Blok for his help with the ATCA
observations. BJP, PMK, and JSG are pleased to acknowledge support
for this work through NASA funds supplied by the Space Telescope
Science Institute.  This research has made use of the NASA/IPAC
Extragalactic Database (NED), which is operated by the Jet
Propulsion Laboratory, California Institute of Technology, under
contract with the National Aeronautics and Space Administration.

\end{document}